\documentclass[aps,twocolumn,prl,superscriptaddress,nofootinbib,amsmath,amssymb,floats,floatfix,english]{revtex4-2}
\usepackage[utf8]{inputenc}
\usepackage[T1]{fontenc}
\usepackage{amsmath}
\usepackage{amssymb}
\usepackage{ifthen}
\usepackage{graphicx}
\usepackage{times}
\usepackage{babel}
\usepackage{color}
\usepackage{bm}
\usepackage[normalem]{ulem}
\usepackage{float}
\usepackage{placeins}
\usepackage{chngcntr}
\usepackage{enumitem}
\usepackage{multirow}
\usepackage{comment}
\usepackage{adjustbox}

\newcommand{\HRR}{1}
\newcommand{\ifhrr}[2]{\ifthenelse{\equal{\HRR}{1}}{{#1}}{{#2}}}
\usepackage[colorlinks=true,citecolor=blue,linkcolor=black,urlcolor=blue]{hyperref}

\newcommand{\ARXIV}{1}
\newcommand{\ifarxiv}[2]{\ifthenelse{\equal{\ARXIV}{1}}{{#1}}{{#2}}}

\newcommand{\ifis}[2]{
\ifthenelse{\equal{#1}{}}{}{#2}
}

\def\be{\begin{equation}}
\def\ee{\end{equation}}
\def\bea{\begin{eqnarray}}
\def\eea{\end{eqnarray}}
\def\bsu{\begin{subequations}}
\def\esu{\end{subequations}}
\def\bi{\begin{itemize}}
\def\ei{\end{itemize}}

\newcommand{\op}[1]{\widehat{#1}}
\newcommand{\dagop}[1]{\widehat{#1}^{\dagger}}

\newcommand{\mc}[1]{{\mathcal{#1}}}
\newcommand{\wt}[1]{{\widetilde{#1}}}

\newcommand{\nonu}{\nonumber}

\newlength{\templength}

\newenvironment{eq}{\begin{equation}}{\end{equation}}

\newcommand{\eqn}[1]{(\ref{#1})}

\renewcommand{\eq}[2]{\begin{equation}\label{#1}#2\end{equation}}
\newcommand{\eqs}[2]{\begin{subequations}\label{#1}\begin{eqnarray}#2\end{eqnarray}\end{subequations}}
\newcommand{\eqa}[2]{\begin{eqnarray}\label{#1}#2\end{eqnarray}}

\usepackage{lmodern}

\setlist[itemize]{nosep}
\setlist[enumerate]{nosep,nolistsep}

\newcommand{\ve}{\varepsilon}

\newcommand{\tick}{\checkmark}

\begin{document}


\newcommand{\titletext}{
Self-organized cavity bosons beyond the adiabatic elimination approximation}
\title{\titletext}

\author{Giuliano Orso}
\email{giuliano.orso@u-paris.fr}
\affiliation{Universit\'e Paris Cit\'e, Laboratoire Mat\'eriaux et Ph\'enom\`enes Quantiques (MPQ), CNRS, F-75013, Paris, France}

\author{Jakub Zakrzewski}
\email{jakub.zakrzewski@uj.edu.pl}
\affiliation{Institute of Theoretical Physics, Jagiellonian University in Krakow, ul. Łojasiewicza 11, 30-348 Kraków, Poland}
\affiliation{Mark Kac Complex Systems Research Center, Jagiellonian University in Krakow, Łojasiewicza 11, 30-348 Kraków, Poland}

\author{Piotr Deuar}
\email{deuar@ifpan.edu.pl}
\affiliation{Institute of Physics, Polish Academy of Sciences, Aleja Lotnik\'ow
32/46, 02-668 Warsaw, Poland}

\begin{abstract}
 The long-time behavior of weakly interacting bosons moving in a two-dimensional optical lattice and coupled to a lossy cavity is investigated numerically via the truncated Wigner method, which
 allows us to take into full account the dynamics of the cavity mode,
quantum fluctuations, cavity-boson correlations, and self-organization of individual runs. 
We first compare our results for small systems with quasi-exact calculations based on quantum trajectories, finding a remarkably good agreement for experimentally relevant boson fillings that improves further with system size. 
For large systems, we observe metastability at very long times and superfluid quasi-long
range order, in sharp contrast with the true long range order found in the ground state of the approximate Bose-Hubbard model with extended interactions, obtained by adiabatically
eliminating the cavity field.
As the strength of the light-matter coupling increases, the system first becomes supersolid at the Dicke superradiant transition and then turns into a charge-density wave via the Berezinskii-Kosterlitz-Thouless mechanism. The two phase transitions are characterized via an accurate finite-size scaling analysis. 
\end{abstract}

\maketitle




\emph{Introduction.} 
Recent experiments explore
the regime 
of strong coupling of quantum matter and quantum light on a variety of platforms, ranging from electronic systems coupled
to THz cavities \cite{Diaz:RMP2019} to atomic Bose
gases coupled to a dissipative cavity \cite{Ritsch:RMP2013,Mivehvar21}, where
 self-organized density patterns have 
been observed \cite{Landig2016,Dreon2022} by tuning the system across the Dicke superradiant transition. Depending on the strength of the atom-cavity coupling, the superfluid (SF) order can either coexist with the charge order, leading to the 
celebrated lattice supersolid (SS) phase,  or disappear, leaving behind 
a charge density wave (CDW) insulator.

Most theoretical works have concentrated on the so-called ``bad cavity'' regime: 
If the  detuning of the cavity $\delta$ and the photon leak rate $\kappa$ are large compared to the characteristic energy scales of the gas, namely the tunneling rate $J$ and the interaction strength $U$, then the cavity mode can be adiabatically eliminated resulting in an extended non-dissipative Bose-Hubbard (EBH) model, characterized by  infinite-range instantaneous interactions. 
The ground state phase diagram of the EBH model obtained using Gutzwiller mean field theories \cite{Droga:PRA2016,Sundar:PRA2016,Himbert:PRA2019}, quantum Monte Carlo \cite{Flottat:PRB2017} and  slave-boson \cite{Sharma:PRL2022} techniques,  predicts that
the SF and SS phases are characterized by a true long-range order 
and that the SS-CDW phase transition is 
of second order \cite{Droga:PRA2016,Sundar:PRA2016,Himbert:PRA2019,Flottat:PRB2017}. Additional interesting phases are found for one dimensional (1D) EBH cavity model \cite{Chanda21,Chanda22}.

For a quantitative comparison with ongoing experiments, a compelling question to address is how the  long-time 
state of the 
full boson-cavity model compares with its simplified 
adiabatic approximation, namely the ground state of the EBH model.
A pioneering step was taken  in \cite{Halati:PRR2020,Halati:PRL2020} using tensor network simulations for a 1D system. Follow on work
suggested \cite{Bezvershenko:PRL2021} that cavity bosons in the steady state  can be described as an equilibrium thermal state of the EBH model,  whose effective temperature is obtained self-consistently, by including perturbatively the effect of quantum fluctuations around the mean field solution. However, this requires computing the retarded
dynamical susceptibility in the EBH model as a function of the frequency, which is challenging for the 2D systems relevant for ongoing experiments \cite{Landig2016,Dreon2022}. 

Importantly, the atom-cavity system differs markedly from photon-polariton systems \cite{DagvadorjPRX2015} and some other non-adiabatic coupled-cavity systems investigated recently \cite{Kelly21,Kelly22,Hosseinabadi23a,Hosseinabadi23b} in that here only the single cavity mode is dissipative, while the overwhelming majority of modes are conservative. Therefore the atoms experience dissipative effects only indirectly, which might 
lead to a very slow relaxation towards the steady state, and make it out-of-reach for 
1D quasi-exact simulations \cite{Halati:PRR2020,Halati:PRL2020}. 

In this Letter we study large 2D systems  through the use of the truncated Wigner (TW) approach \cite{Steel98,Sinatra02,Norrie05,Wouters09,FINESS-Book-Ruostekoski,Chiocchetta14,Lee14,Deuar21}, which keeps full account of the cavity degree of freedom and
naturally describes the bosons with a general mixed state. 
It also allows us to reach long times and compute with a good accuracy
correlation functions and distribution functions
that are crucial to unveil the order of phase transitions and to identify the associated critical points.

\emph{Model and Method.}  The Hamiltonian describing the system can be written as $\op{H} = \op{H}_b + \op{H}_c + \op{H}_{bc}$,
where 
\begin{equation}\label{Hb}
    \op{H}_b= -J \sum_{\langle i,j\rangle} \dagop{b}_i\op{b}_{j} + \frac{U}{2}\sum_j \op{n}_j(\op{n}_j-1) 
\end{equation}
is the Bose-Hubbard model describing the matter component, with $\dagop{b}_j, \op{b}_{j}$  the local  field operators and $\op{n}_j=\dagop{b}_j\op{b}_j$ the local density operator at each lattice site $j$. 
We assume that the cavity contains only one relevant mode of energy $\omega_c$, so that its bare Hamiltonian in the rotating wave approximation is 
 $\op{H}_c = \delta\,\dagop{a}\op{a}$  with $\dagop{a},\op{a}$ being the photon field operators and $\delta=\omega_c-\omega_p$ the detuning between the cavity and the pump frequencies (hereafter we use $\hbar=1$ units).
 The coupling between the bosons and the cavity is described by the Hamiltonian
\begin{equation}
 \op{H}_{bc} = -\frac{\Omega}{\sqrt N}\left(\op{a}+\dagop{a}\right)\op{\Delta},	\label{Hac}       
\end{equation}
where $\Omega$ is the  (re-scaled) Rabi frequency, $N=\langle\sum_j \op{n}_j\rangle$ is the conserved total 
number of bosons, and $\op{\Delta} = \sum_j f_j \op{n}_j$ is a collective operator in which $f_j$ is the overlap of the cavity mode with the wavefunction amplitude in the $j$th lattice site taking, typically, a staggered ``checkerboard'' form \cite{Landig2016}; we use $f_j=(-1)^{j_x+j_y}$ here.  
The single shot values of $\op{\Delta}$ tend to $\pm N$ when atoms are completely arranged in even/odd lattice sites, so it can be considered as a collective magnetization order parameter for even/odd arrangement. The dissipative dynamics of the hybrid system with photon leak rate $\kappa$ is described by the master equation 
\eq{master}{
\frac{d\op{\rho}}{dt} = -i\left[\op{H},\op{\rho}\right] +\kappa \left[2\op{a}\op{\rho}\dagop{a}-\dagop{a}\op{a}\op{\rho}-\op{\rho}\dagop{a}\op{a}\right].
}

In the TW approach  the field operators $\op{a}, \op{b}_{j}$  become represented by an ensemble of complex variables $\alpha, \beta_j$ obeying the
stochastic differential equations \cite{supp}
\begin{eqnarray}
i d\alpha &\!\!=& \!\!\Big[(\delta -i \kappa)\alpha -\frac{\Omega}{\sqrt N}\sum_j f_j\left(|\beta_j|^2-\tfrac{1}{2}\right)\Big]dt + i\sqrt{\kappa} dW_c \nonumber\\
id\beta_j  &=&  \Big[U\left(|\beta_j|^2-1\right)\beta_j 
-\frac{\Omega}{\sqrt N}(\alpha+\alpha^*)f_j\beta_j  \nonumber \\ 
&&-J \sum_{r \in X(j)} \beta_{r}\Big]dt
\label{Eq:TW}
\end{eqnarray}
 where  the set $X(j)$ contains the indices of all the 
 neighboring sites of the site of index $j$ (there are four nearest neighbors in our square $L\times L$ lattice with periodic boundary conditions) and $dW_c$ is a complex-valued Wiener noise term with $\langle |dW_c|^2\rangle=dt$.
 Single trajectories of \eqn{Eq:TW} with a given noise realization correspond closely to experimental runs \cite{Blakie08,Lee14,Lewis-Swan16}, while quantum expectation values are obtained by averaging the operator in symmetric form
 over many trajectories (numbering 240-2400 here).  
 For example, for the boson local density $\langle \op{n}_j \rangle=\langle(\dagop{b}_j\op{b}_{j}+\op{b}_{j}\dagop{b}_j-1)/2\rangle =\langle|\beta_j|^2-\tfrac{1}{2}\rangle$.
 Remarkably the approach scales only linearly with the number of sites, giving access to large systems (the $64\times64$ lattices treated here are still far away from any significant limitations \cite{Deuar21}), 
and to the very long evolution times needed to reach the steady state,  because it is not inherently limited by the entanglement growth of  tensor network methods 
\cite{Halati:PRL2020,Halati:PRR2020,Bezvershenko:PRL2021}. 

\begin{figure}
\hspace*{-0.3cm}\includegraphics*[width=\columnwidth]{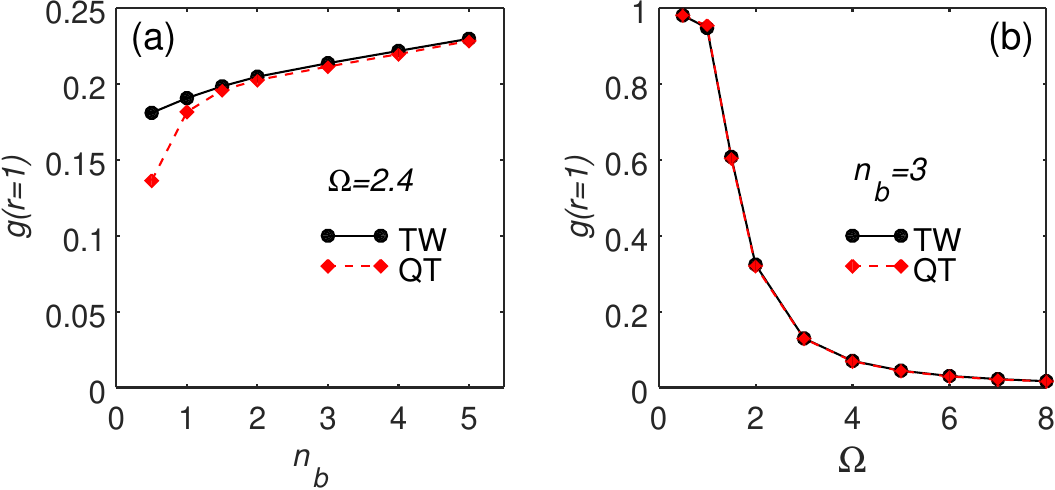}\\
\includegraphics*[width=\columnwidth]{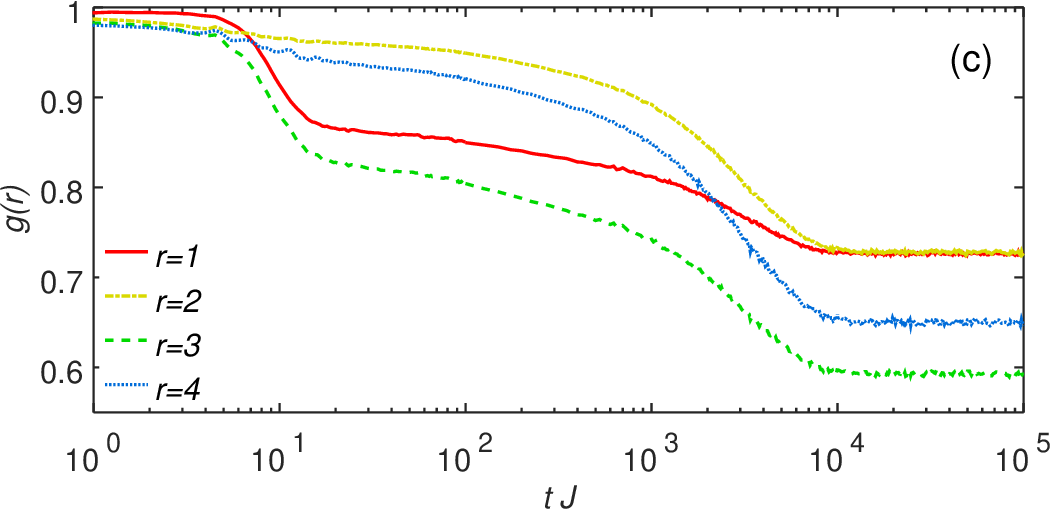}
\caption{Coherence and metastability of $g(r)$, with benchmarking on small systems. Panels (a-b) compare TW results (black circles) with quasi-exact quantum trajectories calculations (red diamonds) in $2\times 1$ lattices, for the initial 
wave-function $|\textrm{vac}\rangle  \otimes |\beta_1\rangle \otimes  |\beta_2\rangle$,
with vacuum cavity and 
bosons at each site in a coherent state, 
with $\beta_i=\sqrt{N/2}$.  Steady state values of the superfluid phase correlation \eqn{Eq:g} at distance $r=1$ are plotted on (a) the boson filling for fixed $\Omega=2.4J$  and (b) the Rabi frequency for fixed $n_b=3$. Panel (c) shows superfluid phase correlation versus evolution time for a $16\times 16$ lattice with $\Omega=2.4J$ and $n_b=5$; Curves red to blue show four successively increasing distances $r=1$ to $4$. The other model parameters are $\delta=2J$, $\kappa=0.5J$, $U=0.2J$.
Note the metastable plateau at times $tJ\sim\mc{O}(100)$. 
}
\label{Fig:longtime}
\end{figure}

In the following we take $\kappa=J/2$, $\delta=2J$ and $U=0.2J$,
corresponding to a regime in which the energy scales of the cavity and of the matter components are of the same order
so that fluctuation, retardation or dissipation effects can play a significant role.
Evolution times are  $tJ=10^4$ (i.e. about 200 times longer than in \cite{Halati:PRL2020} or \cite{Lee14}) unless otherwise noted, starting from an initial state with atoms in a coherent state
and vacuum in the cavity photon mode. Deviations from the adiabatic elimination estimate are indeed observed very clearly for these parameters\cite{supp}, being $\mc{O}(1)$ for  $\op{a}$.
We characterize the nature of the different phases by computing the superfluid  (phase) correlation function
\begin{equation}\label{Eq:g}
g(r) = \frac{1}{N}\sum_j \textrm{Re}(\langle\dagop{b}_j\op{b}_{j+r}\rangle),
\end{equation}
where the index $j+r$ is modulo $L$  to conform to the periodic boundary conditions.

To assess the accuracy of the TW approach, we first focus on a two-site system and compare our long-time predictions for $g(r=1)$ with quasi-exact
calculations using
the quantum trajectories
method \cite{Dalibard:PRL1002,Dum:PRA1992}, 
with the initial boson wave-function at each site being a coherent state. The results are displayed in Fig.~\ref{Fig:longtime}a as 
a function of the boson filling for $\Omega=2.4J$, 
revealing a remarkably good agreement for  $n_b\gtrsim 1.5$, with a relative error around one percent. Importantly, our approach is  well suited for a quantitative comparison with experiments \cite{Landig2016}, where the reported filling
at the trap center is between $2$ and $3$. 
In Fig.~\ref{Fig:longtime}b we display $g(1)$ vs $\Omega$
for fixed $n_b=3$, showing that
the TW approach remains very
accurate as the coupling to the cavity increases and the 
coherence between the two sites is progressively lost. Numerical and analytical demonstration of further improvement of accuracy with lattice size and discussion of the TW 
validity criteria are in \cite{supp}.

\emph{Metastability and self-organization.} Hereafter we consider large 2D lattices and  fix $n_b=5$ for definiteness.
Metastable dynamics is observed for many $\Omega$ values, especially in the range $2\lesssim\Omega/J\lesssim5$.
A typical example is shown in Fig.~\ref{Fig:longtime}c where the time dependence of the correlation $g$
is plotted for several distances $r$. Here, a quasi stationary state appears at the times $tJ\sim\mc{O}(20)$ reached by previous studies, while true convergence occurs only at $tJ \approx 10^4$. 

\begin{figure}[tb]
\centering
\includegraphics[width=\columnwidth]{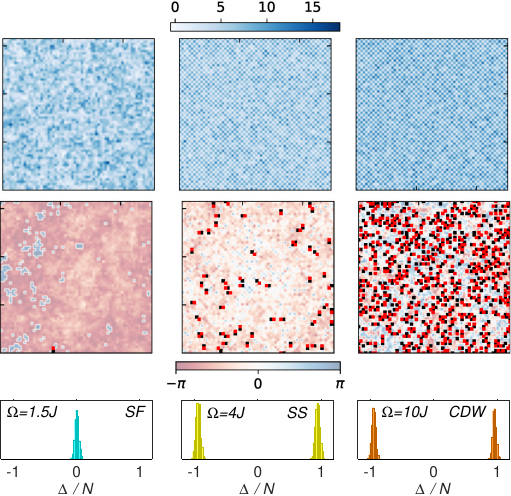}
\caption{Long-time density (top) and phase (middle) distributions of bosons of a given TW trajectory  on a $64\times 64$ lattice at $tJ=10^4$. 
The 
columns correspond to three increasing values of the coupling Rabi frequency  $\Omega=1.5J$ (left), $4J$ (center), and $10J$ (right), where the system is in the SF, SS, and CDW 
phases, respectively.
Other parameters 
are as in Fig.~\ref{Fig:longtime}c.
The  square symbols in the middle row signal the presence of a vortex (black) or an antivortex (red) in the plaquette. 
Bottom: distribution of the density wave order parameter $\Delta$ over an 
ensemble of 800 trajectories on $16\times16$ lattices.
}
\label{Fig:vortices}
\end{figure}

As the coupling $\Omega$ is increased, three phases are encompassed, with typical observables in single shots shown in  Fig.~\ref{Fig:vortices}.
For small values of $\Omega$, the density (top)
shows strong site-to-site fluctuations, while the phase (lower) is essentially uniform, corresponding to a SF phase. As $\Omega$ 
increases, 
the density distribution becomes a checkerboard, 
signaling the appearance of a density wave order and antibunching correlations between neighboring sites. The actual order is self-organized spontaneously, as evidenced by  the $\Delta/N$ distribution peaked at $\pm1$, corresponding to the occupation of even or odd  sites. At the same time,
the phase distribution shows sizable fluctuations, along with the formation of scattered vortex-antivortex pairs, that we identify by calculating the phase changes accumulated in a plaquette of nearest neighbors. 
Their binding implies 
that the system possesses both SF and charge order, the hallmark of the SS phase. A further increase of the  Rabi frequency (right) leads to a more pronounced staggering of the 
density 
that is accompanied by a massive proliferation of vortices and antivortices, which act to destroy the SF order. We identify this normal phase as a CDW insulator and argue that the SS-CDW phase transition should obey the BKT scenario, according to which superfluidity in 2D systems with $U(1)$ symmetry is lost due to the unbinding of the topological excitations.

\begin{figure}
\includegraphics[width=\columnwidth,clip]{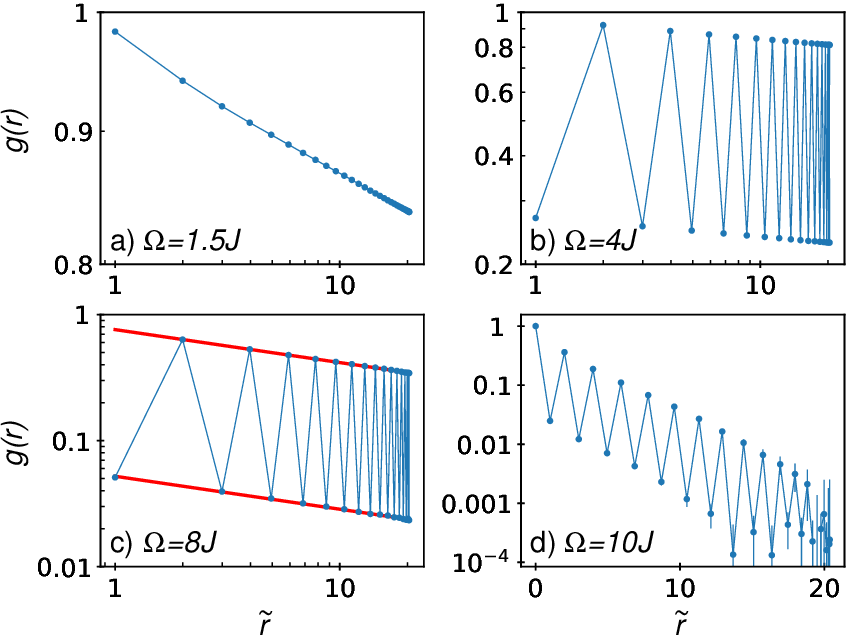}
\caption{
Long-time behavior of the 1st order correlation \eqn{Eq:g}
at $tJ=10^4$, versus the lattice distance $r=|j_x-i_x|$ along one direction, averaged 
over 240 trajectories for a $64\times64$ lattice. 
More precisely, $\wt{r}=(L/\pi)\sin(\pi r/L)$ is the chord distance.
 Notice the change from logarithmic to linear scale in the horizontal axis of panel d), emphasizing the transition from power-law decay QLRO (a-c), 
to exponential decay (d) in the proximity of the BKT critical point (see text).
We extract the exponents $\eta$ from fits (red solid lines). 
}
\label{Fig:gcorrelation}
\end{figure}

\emph{Quasi long-range order (QLRO).}
To characterize the atom 
superfluidity in the long-time regime, 
 we inspect the spatial dependence of the correlation \eqn{Eq:g} in Fig.~\ref{Fig:gcorrelation}.
In the SF and SS phases it displays a power-law decay of the form $g(r)=(a+ (-1)^r b\,)/\tilde r ^{\eta}$, where $b$ is nonzero in the presence of solid order and 
$\wt{r}=(L/\pi)\sin(\pi r/L)$ is the ``chord'' distance, 
which accounts to leading order for the periodic boundary conditions  \cite{Gerster:NJP2016}.
For $\Omega=8J$ (c), where the system is close to the SS-CDW transition (see below), a fit to the data, with $a, b, \eta$ as free fitting parameters, yields  $\eta=0.261(5)$, which is slightly larger than 
the critical anomalous exponent $\eta_c^{XY}=1/4$, holding for the XY universality class at equilibrium.  
 We stress that the cavity-induced QLRO observed here contrasts with the true long-range SF order observed in previous ground state calculations of the 2D EBH model, reflecting the shortcomings of the cavity adiabatic elimination. 
 In the CDW phase (d) the  function $g(r)$ decays exponentially, consistently with the BKT transition to an insulator.

\begin{figure}
\includegraphics[width=\columnwidth]{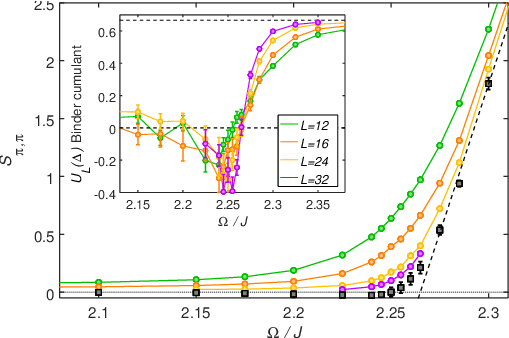}
\caption{Main plot: static structure factor $S_{\pi,\pi}(L)$ versus $\Omega$ 
across the superfluid-supersolid phase transition, showing finite size scaling. Black points represent the $L\to\infty$ thermodynamic limit  $A(\Omega)$, according to \eqn{SpiL2}. The dashed black line is a linear fit of $A(\Omega)$ in the above threshold phase, giving the Dicke transition point as $\Omega_s=2.265(5)$. 
The inset shows the Binder cumulant $U_L$ for the ``magnetization'' $\Delta/N$ across the transition. Here the crossing of $U_L=0$ occurs at $\Omega_s=2.265(2)$.
Data at $tJ=10^4$, 800 trajectories. 
}
\label{Fig:Dicke}
\end{figure}

\emph{Identification of critical points.} We next employ the finite-size scaling approach to pinpoint the critical values $\Omega_s$ and $\Omega_{c}$  where the two phase transitions occur and characterize their nature. 
At $\Omega=\Omega_s$ the  system undergoes the Dicke superradiant transition, where the cavity mode becomes macroscopically occupied and, concomitantly, the solid order appears \cite{Dicke54,HEPP1973360}.
 We characterize the latter via the static structure factor 
$S_q= \sum_{ij} e^{i q(i-j)} (\langle \op{n}_i \op{n}_j\rangle -\langle \op{n}_i\rangle \langle \op{n}_j \rangle)/L^4$,
with $q=(\pi,\pi)$.
 Fig.~\ref{Fig:Dicke}  displays the $\Omega$-dependence  of $S_{\pi,\pi}$ for increasing system sizes. The cavity occupation shows the same features, as shown in \cite{supp}.  
 In the SF phase the structure factor scales to zero as $1/L^2$, while in the SS phase it converges to a finite value. From a finite-size scaling analysis according to \cite{supp}
 \eq{SpiL2}{
 S_{\pi,\pi}(\Omega,L)=A(\Omega)+\frac{C(\Omega)}{L^2}
 }
 and then following the visible linear trend of $A(\Omega)\propto \Omega-\Omega_s$   we obtain $\Omega_s=2.265(5)$. 
The linear in $\Omega$ trend 
close to the critical point, also observed for 1D systems \cite{Halati:PRL2020}, confirms that the Dicke superradiant transition 
can be described with Landau (mean-field) theory \cite{Kirton18}. 

The TW method also allows us to calculate the Binder cumulant $U_L(\Delta)=1-\langle\Delta^4\rangle/3\langle\Delta^2\rangle^2$ for the effective magnetization $\Delta$. A quantity well suited for locating phase transitions \cite{Binder84} but ruled out {in the} mean field or bad cavity approximations, and therefore 
not investigated to date. It is shown in Fig.~\ref{Fig:Dicke}(inset) and provides the more accurate value $\Omega_s=2.265(2)$ for the critical point at which the jump from $U_L=0$ to $U_L=2/3$ appears in the $L\to\infty$ limit. Before the transition there is 
an interesting regime in which $U_L<0$, that appears to stem from the appearance of fatter tails in the probability distribution of $\Delta$ as a precursor of the phase transition  \cite{supp}.

\begin{figure}
\vspace*{-1em}
\includegraphics[width=1\columnwidth]{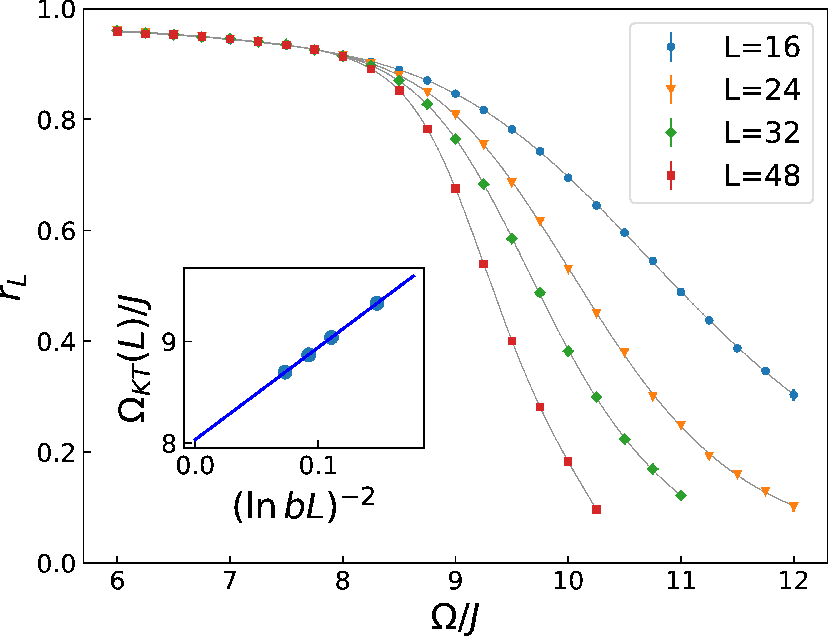}
\vspace*{-1em}
\caption{Main panel: correlation ratio $r_L=g(L/2)/g(L/4)$ at 
$tJ=10^4$, $2400$ trajectories, as a function of $\Omega$
for different system sizes (data symbols). The solid lines are fits to the data curves with Padé approximants $p/q$,   where $p$ and $q$ have polynomial degrees $n_p=3$ and $n_q=4$.
Inset: Identification of the BKT critical point based on
$R=0.8$ (see text).}
\label{Fig:rL_vs_Omega}
\end{figure}

We now turn to the BKT transition. For periodic boundary conditions, a useful indicator is provided by the correlation ratio~\cite{Tomita:PRB2002} $r_L=g(L/2)/g(L/4)$,  
In Fig.~\ref{Fig:rL_vs_Omega} we plot the long-time behavior of $r_L$ 
against $\Omega$ for increasing system sizes. 
For $\Omega \lesssim 8J$, the correlation ratio is essentially size independent, reflecting the QLRO. In contrast, for $\Omega \gtrsim 8J$ the different curves spray out, consistent with the fact that the correlation ratio must scale to zero in the CDW phase, suggesting that $\Omega_c \approx 8J$.  
 For a more accurate estimate, we use the fact that the correlation ratio is described by a single parameter scaling 
function, $r_L=f (L/\xi)$, where $\xi=\exp(c/\sqrt {\Omega/\Omega_{c}-1})$
is the correlation length
in the normal phase $(\Omega>\Omega_c)$
with $c>0$ being a model-dependent constant. 
Following~\cite{Tomita:PRB2002}, we define $\Omega_{KT}(L)$  as the 
value of $\Omega$ at which $r_L=R$.
From $\xi$ in the scaling regime $\Omega_{KT}(L)$  satisfies $L\exp(-c/\sqrt u)=1/b$, where $u=\Omega_{KT}(L)/\Omega_c-1$ and $b$ is related to $R$ via $f(1/b)=R$, so that~\cite{Tomita:PRB2002} 
\begin{equation}\label{Eq:FSS}
 \Omega_{KT}(L)=\Omega_c +\frac{c^2 \Omega_c}{\ln^2 bL}.   
\end{equation}
We use Eq.~(\ref{Eq:FSS}) to fit our data, with $\Omega_c, c$ and $b$ being fitting parameters and $R=0.8$. The fit is 
displayed in the inset of Fig.~\ref{Fig:rL_vs_Omega}, 
confirming that the
linear dependence is well satisfied, 
with the intercept  $\Omega_c/J=8.03(4)$. Different choices for $R$, with $0.6 \lesssim R < 0.8$, yield
$\Omega_c$ consistent within error with the above.

\emph{Conclusions.}
We have demonstrated that the TW method allows
one to accurately investigate 
the long-time 
properties of  large 2D  boson-cavity lattices, without being hindered by the bad cavity approximation, by the entanglement growth, or by 
specific assumptions on the nature of the atomic mixed state.  
We confirm that
the system is
prone to a very slow evolution that can be missed by previous time-limited approaches, which are only able to reach the initial plateau.  Such long timescales  
are nevertheless   
relevant for ongoing experiments \cite{Landig2016} that typically deal with times $tJ\sim10^4-10^5$. We have unveiled that the SF and  SS phases exhibit QLRO, as compared to the true long-range order suggested by 
the EBH model at zero temperature.  The QLRO in turn
disappears at the SS-CDW phase boundary according to the BKT scenario. 
Our results open a new avenue for a direct and quantitative comparison between theory and ongoing experiments on boson-cavity systems.

Finally, we note that research just published finds self-organization phenomena not captured by an adiabatic elimination approach also for fermionic atoms in a dissipative cavity\cite{Tolle25}.

\ifthenelse{\equal{\ARXIV}{0}}{
\acknowledgments
}{}
We are grateful to I. Carusotto, C. Ciuti, F. Hebert, S.B. Jager, King Lun Ng, and T. Roscilde  for helpful explanations and discussions. 
This research was co-financed by the Polish National Agency for Academic Exchange,  grant No. BPN/BFR/2022/1/00027 and by the HPC Polonium grant No. 49317RK. 
This work was granted access to the HPC resources of TGCC under the allocation AD010513635 and AD010513635R1 supplied by GENCI (Grand Equipement National de Calcul Intensif). This research was also funded by National Science Centre (Poland) under the OPUS call within the WEAVE programme 2021/43/I/ST3/01142 (J.Z.).

\renewcommand{\bibsection}{}   
\bibliography{artnew}

\clearpage

\renewcommand{\thefigure}{S\arabic{figure}}
\setcounter{figure}{0}
\renewcommand{\thetable}{S\arabic{table}}
\setcounter{table}{0}
\renewcommand{\thesection}{S\arabic{section}}
\setcounter{section}{0}
\renewcommand{\thesubsection}{\Alph{subsection}}
\renewcommand{\theequation}{S\arabic{equation}}
\setcounter{equation}{0}
\renewcommand{\thepage}{s\arabic{page}}
\setcounter{page}{1}
\setcounter{footnote}{0}

\begin{widetext}
\begin{center}
{\bf Supplementary material for: \\  
\titletext}
\normalsize
\vspace*{1em}

{Giuliano Orso$^1$, Jakub Zakrzewski$^{2,3}$, and Piotr Deuar$^4$}\\[1em]
{\small
\textit{$^{1}$Universit\'e Paris Cit\'e, Laboratoire Mat\'eeriaux et Ph\'eenom\`enes Quantiques (MPQ), CNRS, F-75013, Paris, France\\
$^2$ Institute of Theoretical Physics, Jagiellonian University in Krakow, ul. Łojasiewicza 11, 30-348 Kraków, Poland\\
$^3$ Mark Kac Complex Systems Research Center, Jagiellonian University in Krakow, Łojasiewicza 11, 30-348 Kraków, Poland\\
$^4$ Institute of Physics, Polish Academy of Sciences, Al. Lotnik\'ow 32/46, 02-668 Warsaw, Poland}
}
\vspace*{1em}
\end{center}
\end{widetext}
\tableofcontents 

\vspace*{2em}
\noindent\textsl{\underline{Note:} Citation numbers in square brackets refer to the bibliography in the main paper. }

\maketitle

\section{S1. Truncated Wigner Methodology}
\label{S-TW}

\subsection{A. Overview of the method and its derivation}

The ``truncated'' Wigner method is an adaptation of the widely known Wigner phase-space quasiprobability distribution $W(x,p)$ for representing the state of a quantum mode, made to allow its application to very large many-mode systems. It is quite widely used e.g. in ultracold atom and quantum optics systems \cite{Steel98,Drummond01a,Sinatra02,Norrie05,FINESS-Book-Ruostekoski} and also has been applied to polariton and open systems of the Bose-Hubbard type \cite{Wouters09,Chiocchetta14,Deuar21}, as well as cavity self-organized systems similar to ours in one dimension \cite{Lee14,Lee15}.

For $M$ sites, instead of studying the quasiprobability distribution $W(\vec{x},\vec{p})$ as a distribution in a $2M$-dimensional space, values of the amplitude vector $\vec{\alpha}=\vec{x}+i\vec{p}$ are sampled according to the distribution $W(\vec{\alpha})$. Here amplitudes $\alpha_j=x_j+ip_j$, corresponding to space and momentum operators $\op{x}_j=(\op{a}_j+\dagop{a}_j)/2$ and $\op{p}_j=(\op{a}_j-\dagop{a}_j)/2i$ in an abstract ``harmonic oscillator space'' for each site $j$ are similar (though not equal) to coherent state amplitudes.
These amplitudes then obey stochastic differential equations with noise amplitudes appropriate to represent the dissipative quantum mechanical evolution. Appropriate averages over the ensemble of $S$ samples or ``trajectories'' correspond then better and better to the quantum state as $S\to\infty$.

More precisely, 
the density matrix $\op{\rho}$ can be expanded in terms of the Wigner quasiprobability distribution as 
\eq{Wrho}{
\op{\rho} = \int d\vec{\alpha}\,W(\vec{\alpha})\op{\Lambda}_W(\vec{\alpha}),
}
where $\op{\Lambda}_W$ is an operator kernel \cite{Cahill69b,Deuar21b} dependent on the set of complex phase space amplitudes $\vec{\alpha}=\{\alpha,\beta_j\}$, corresponding to the 2nd quantized operators $\op{a}, \op{b}_j$ for the photon mode and boson sites, respectively. 
With the correct Wigner representation kernel, operators obey the differential identities:
\begin{subequations}\eqa{ident}{
\op{a}\op{\Lambda}_W &=& \left[\alpha -\frac{1}{2}\frac{\partial}{\partial\alpha^*}\right]\op{\Lambda}_W\\
\dagop{a}\op{\Lambda}_W &=& \left[\alpha^* +\frac{1}{2}\frac{\partial}{\partial\alpha}\right]\op{\Lambda}_W\\
\op{b}_j\op{\Lambda}_W &=& \left[\beta_j -\frac{1}{2}\frac{\partial}{\partial\beta_j^*}\right]\op{\Lambda}_W\\
\op{b}^{\dagger}_j\op{\Lambda}_W &=& \left[\beta_j^* +\frac{1}{2}\frac{\partial}{\partial\beta_j}\right]\op{\Lambda}_W,
}
\end{subequations}
and their adjoints. 
With their help, using standard phase-space representation procedure, we can rewrite the master equation \ifarxiv{\eqn{master}}{(3)}
as an integral equation 
\eqa{inteq}{
\lefteqn{\int d\vec{\alpha}\,\frac{dW}{dt}\op{\Lambda}_W}&&\\
&&= \int d\vec{\alpha}\,W \left[\sum_k A_k\frac{\partial}{\partial\alpha_k}+\sum_{kk'}\frac{D_{kk'}}{2}\frac{\partial^2}{\partial\alpha_k\partial\alpha_{k'}} + \dots\right]\op{\Lambda}_W\nonu
}
where $\alpha_k$ label all variables in $\vec{\alpha}$ and ``$\dots$'' indicate higher order partial derivatives (up to 3rd order in our case, as it turns out). 
Integrating by parts and assuming vanishing boundary terms at $|\alpha|,|\beta|_j\to\infty$, this gives $\int d\vec{\alpha}\,\op{\Lambda}_W\frac{dW}{dt} = \int \op{\Lambda}_W\left[\circ\circ\circ\right]W$, which has at least one solution in the form of a Fokker-Planck equation (FPE) with additional third order terms:
\eqa{fpe}{
\frac{dW}{dt} &=& \left[\circ\circ\circ\right]=\Bigg[\sum_k \frac{\partial}{\partial\alpha_k}(-A_k)+\sum_{kk'}\frac{\partial^2}{\partial\alpha_k\partial\alpha_{k'}}\frac{D_{kk'}}{2} \nonu\\
&&\qquad- \sum_{kk'k"}\frac{\partial^3}{\partial\alpha_k\partial\alpha_{k'}\partial\alpha_{k"}}T_{kk'k"} \Bigg]W.
}
In our particular case, in hopefully self-evident notation, the nonzero elements are
\eqa{ADT}{
A_{\alpha} &=& -i\delta\alpha -\kappa\alpha +i\frac{\Omega}{\sqrt{N}}\sum_jf_j\left(|\beta_j|^2-\tfrac{1}{2}\right)\nonu\\
A_{\beta_j} &=& -iU(|\beta_j|^2-1)\beta_j+i\frac{\Omega}{\sqrt{N}}(\alpha+\alpha^*)f_j\beta_j\nonu\\
&&+iJ\sum_{r\in X(j)}\beta_r\nonu\\
D_{\alpha\alpha^*} &=& D_{\alpha^*\alpha} = \kappa\\
T_{\beta_j\beta_j\beta^*_j} &=& i\frac{U\beta_j}{2}\nonu\\
T_{\beta_j\beta_j^*\alpha} &=& -i\frac{\Omega f_j}{4\sqrt{N}},\nonu
}
as well as $A_{\alpha_k^*}=(A_{\alpha_k})^*$ and $T_{\alpha_k^*\alpha_k'^*\alpha_k"^*} = (T_{\alpha_k\alpha_k'\alpha_k"})^*$.

The $A$ and $D$ terms lead to drift and diffusion processes for samples $\vec{\alpha}$ of the distribution.  Although some notable attempts at including third order terms like $T$ numerically have been made \cite{Plimak01,Polkovnikov03,Drummond14}, there are no known well scalable implementations. The standard \textit{truncated Wigner} procedure is to now remove these offending third order terms but remain in a regime where their effect is negligible. Studies have shown that this is justified in ultracold atom systems when some conditions are met, especially that the mean occupation per mode is greater than $\mc{O}(\tfrac{1}{2})$ (single mode occupations need not) \cite{Norrie06}. In our case the mean occupation is $n_b=5$. TW accuracy is looked at in more detail in Sec.~\ifarxiv{S1 E}{\ref{S-TW-scale}}. 

The Fokker-Planck equation \eqn{fpe}, after removal of the $T_{kk'k"}$, corresponds to the following stochastic Langevin equations in the Ito calculus:
\eq{Lang}{
d\alpha_k = A_{\alpha_k} dt + \sum_{l} B_{\alpha_k,l}\, dW_l 
}
where the diffusion is decomposed as $D=BB^T$ using noise matrices $B$ with matrix elements $B_{\alpha_k,l}$. The $dW_l$ are real Wiener increments with $\langle dW_l\rangle=0, \langle dW_ldW_{l'}\rangle=\delta_{ll'}$. In our case the noise matrix is very simple, and one can take
\eq{B}{
B_{\alpha,1} = B_{\alpha^*,1} = \sqrt{\frac{\kappa}{2}}, B_{\alpha,2}=-B_{\alpha^*,2}=i\sqrt{\frac{\kappa}{2}}. 
}
as the only nonzero elements. Substitution of \eqn{ADT} and \eqn{B} into the general form \eqn{Lang} yields the truncated Wigner evolution equations for the system, \ifarxiv{\eqn{Eq:TW}}{(4)}.

While the evolution equations \ifarxiv{\eqn{Eq:TW}}{(4)}
bear a lot of resemblance to the c-field or ``Gross-Pitaevskii'' formulation, obtained by removing the noises $dW_l$, the interpretation of the amplitudes is different. The Wigner distribution of a vacuum state is a Gaussian $W(\alpha) = \frac{2}{\pi}\exp\left[-2|\alpha|^2\right]$ with mean $\langle|\alpha|^2\rangle=\tfrac{1}{2}$, and of a coherent state $\alpha_0$ the same Gaussian centered at $\alpha_0$. Therefore, it includes ``virtual vacuum noise'' that encodes the Heisenberg-limited minimum uncertainty between non-commuting operators such as $\op{a}$ and $\dagop{a}$ via the randomness of the ensemble.

\subsection{B. Initial conditions and observables}
Our initial conditions are vacuum in the cavity photon mode and a condensate (coherent state) with mean occupation $n_b=5$ per site in the atomic lattice.
i.e. \cite{Olsen09}
\eq{ictw}{
\alpha_j(0) =  \frac{\xi}{\sqrt{2}}\qquad;
\qquad
\beta_j(0)=\sqrt{n_b}+\frac{\chi_j}{\sqrt{2}}.
}
Here $\xi$ and $\chi_j$ are independent complex random variables, whose real and imaginary parts are normal random variables with zero mean and standard deviation equal to $1/\sqrt{2}$, so that $\langle|\chi_j|^2\rangle=\langle|\xi|^2\rangle=1$ and other 1st and 2nd order stochastic averages are zero. 
We use a coherent initial state because the Wigner distribution does not describe a Fock state conveniently (though it is possible to do if needed, see \cite{Olsen09}).

In concert with the virtual vacuum fluctuations, physical observables must also be calculated in a more nuanced way than simply replacing $\op{b}_j\to\beta_j$ as one would do with mean field treatments. Expectation values of an observable $\op{O}$ are found by taking the average of its corresponding Weyl symbol (which can be obtained from application of \eqn{ident} and \eqn{Wrho} on Tr[$\op{O}\op{\rho}$]). In particular, 
\eqs{obsTW}{
\langle\dagop{b}_j\op{b}_i\rangle &=& \langle\beta_j^*\beta_i\rangle-\frac{\delta_{ij}}{2},\\
\langle\op{b}_j\rangle &=&\langle\beta_j\rangle\\
\langle\op{n}_j\op{n_i}\rangle &=& \left\langle\left(|\beta_j|^2-\frac{1}{2}\right)\left(|\beta_i|^2-\frac{1}{2}\right)\right\rangle -\frac{\delta_{ij}}{4},\quad
}
and analogous expressions for the photon mode.
\eqn{obsTW} are then used to calculate observables such as occupations, phases, $g(r)$, $S_{\pi,\pi}$, and $\Delta$. 

Single trajectories of a TW simulation are known to correspond to a significant degree to experimental measurements, on the same basis that Gross-Pitaevskii simulations are a good predictor of single experimental runs. Only here, quantum fluctuation effects are additionally taken into account \cite{Blakie08}. TW trajectories correspond well to the classical stochastic measurement trajectories of a continuously monitored system \cite{Lee14}, here monitoring of the leaking cavity photons by the environment.
The correspondence is generally up to the level of minimum Heisenberg-limited uncertainty which is reflected by the virtual noise in the initial conditions. That is -- uncertainty on the order of $\tfrac{1}{2}$ particle on one site, $1/\sqrt{2n_j}$ on the local phase at site $j$, or $\sqrt{L^2}/2$ for density observables on the whole system. As a reflection of this, for example, single trajectories do not capture $\pm1$ modulation of particle number in e.g. squeezed states \cite{Lewis-Swan16}, though ensemble averages remain fine. 
For $\Delta/N$ such uncertainty/washing out comes out of the order of $1/(2Ln_b)$. As we can see, sizeable site occupations such as $n_j\sim n_b$ are expected to lead to a qualitatively correct representation of single runs under most conditions.

On the other hand non-observable quantities such as entanglement entropy or purity of the state are not accessible, except through emulating quantum tomography measurements.

\subsection{C. Simulation details}
\label{S-para-crit}

As per the main text, physical parameters are
$\kappa=J/2$, $\delta=2J$, $U=0.2J$ and $n_b=N/L^2=5$, the mean number of particles per site. In the SF and CDW phases, local mean density can approach $n_j\approx10$, with neighbors $n_{j+1}\approx\mc{O}(1)$. 
The atoms are on a square 2d lattice with periodic boundary conditions, of dimensions $L\times L$. Total evolution times (except for Fig.~2) are $tJ=10^4$, which is also a typical waiting time used in experiments \cite{Landig2016}. For expectation values, 240-2400 trajectories were averaged, depending on the case and the signal to noise ratio.

We have used two kinds of integration scheme, both of which give mutually consistent results.

\noindent(i) the Runge-Kutta 4 method with the time step $dt$ chosen such that the numerical relative error in the conservation of the total boson number $N$ is sufficiently small, $|N-\sum_j \langle \hat n_j\rangle|/N<10^{-4}$ over the evolution time; For the range of parameters explored this yields a time step varying between $10^{-4}J^{-1}$ and $10^{-3}J^{-1}$.

\noindent
(ii) a symmetric split step method in which the terms of $d\beta_j$ in $U$ and $\Omega$ are performed exponentially in real space over $dt$, 
\eq{ssx}{
\beta_j \to \beta_j \exp\left\{
-idt\left[U(|\beta_j|^2-1)-2\frac{\Omega}{\sqrt{N}}{\rm Re}[\alpha]f_j\right]
\right\}
}
while the tunneling term in $J$ is performed in two half-steps of $dt/2$ in k-space (reached by a Fourier transform of $\beta_j$ to $\wt{\beta}_{\vec{k}}$), using the exponential factors 
\eq{ssk}{
\wt{\beta}_{\vec{k}} \to \wt{\beta}_{\vec{k}}\exp\left[-i\left(\frac{dt}{2}\right)2J\sum_{l=1,2}\left(1-\cos\left(\frac{2\pi k_l}{L}\right)\,\right).\right]
}
Here $\vec{k}=[k_1,k_2]$ are the lattice indices in k-space, in dimensions $l=1,2$. Double half-steps help to bring the integration error in local $n_j$ down to $\propto dt^3 t$.
This split-step approach conserves total atom numbers exactly, and requires a $dt\sim10^{-3}J^{-1}$.

\subsection{D. Comparison with quasi-exact numerics}

We assess the accuracy of the TW on $2\times 1$ lattices, for which quasi-exact numerics is possible via the Monte Carlo wave-function method, based on quantum trajectories. The initial wave-function is chosen as $|\textrm{vac}\rangle  \otimes |\beta_1\rangle \otimes  |\beta_2\rangle$,
where $|\textrm{vac}\rangle $ is the vacuum state for the cavity field and $|\beta_i\rangle$ are coherent states for the
bosons at site $i=1$ and $2$, with $\beta_i=\sqrt{N/2}$.
We express the wave-function in the  Fock space (occupation number) representation, by 
introducing two cut-off parameters, corresponding to the maximum cavity photon number $N_{a\textrm{max}}$ and the maximum boson number $n_{b\textrm{max}}$ per site. We increase their values until the numerical results have fully converged. For the data shown in Fig.~1ab of the main text we have
found that this requires (at most)  $N_{a\textrm{max}}=80$ and $n_{b\textrm{max}}=18$.

\begin{figure}[htb]
\includegraphics[width=\columnwidth]{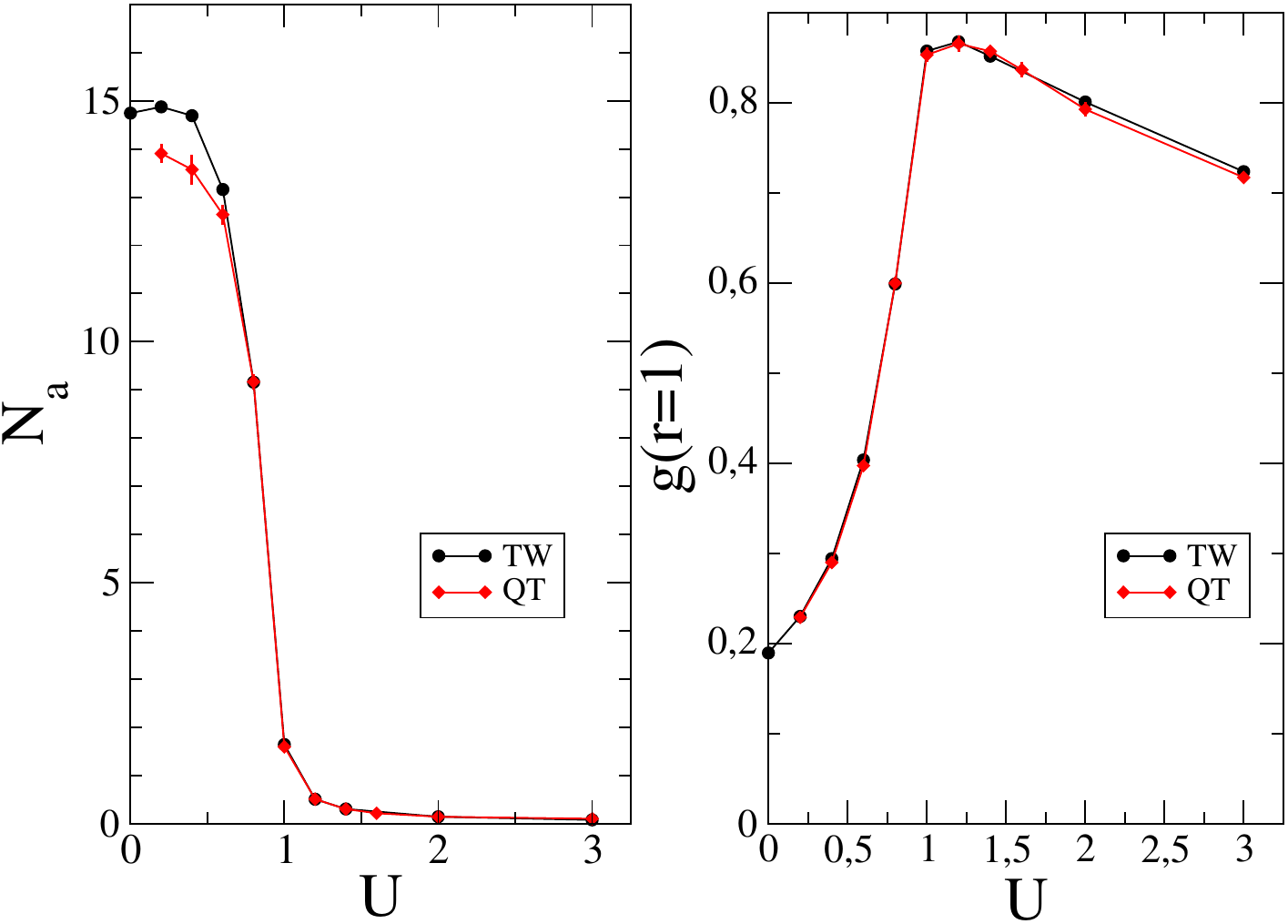}
\caption{
Comparison of cavity occupation $N_a$ (left) and the nearest neighbor value $g(1)$ of the superfluid correlation function (right) vs the interaction strength for a two-site lattice with $n_b=5$ and $\Omega=2.4J$ at $t=400J^{-1}$, in the steady state. The black circles correspond to the TW prediction, while the red diamonds represent the results of the Monte Carlo wavefunction method, both with initial coherent states at each lattice site.
All the other values of the model parameters are as in Fig. 1c of the manuscript.
\label{fig:compareTW-QT}}
\end{figure}

The non-unitary time evolution of the wave-function between successive quantum jumps is calculated via the 4th-order Runge-Kutta 
method. For a better accuracy, we perform two
integrations with time step $dt$ and $dt/2$ and then apply Richardson's extrapolation
formula. The time step used to obtained the
results shown in Fig.~1ab was chosen between $0.01J^{-1}$ and $0.001J^{-1}$ depending on the values of the model parameters. 

For completeness in Fig. \ref{fig:compareTW-QT} we display $g(r=1)$ and the cavity occupation $N_a$
as a function of the Hubbard interaction $U$, 
for filling $n_b=5$. We see that the TW prediction of $g$ remains very close to the quasi-exact result even for $U=3J$ and all the way down to $U=0$. 
For $N_a$, only a small difference is seen at low $U$ values. 
For small $U$, the cavity occupation $N_a$ remains close to its non-interacting value, implying that the discrepancy with the exact result comes mainly from the discarded 3rd order term proportional to the boson-cavity coupling $\Omega$. 
We will see in Fig.~\ref{fig:scalingTW-QT} and the analytic scaling analysis in Sec.~\ifarxiv{S1 E}{\ref{S-TW-scale}} below that any discrepancy drops rapidly as the system size or $n_b$ grow.

\subsection{E. Scaling with system size}
\label{S-TW-scale}

Since this work deals with phenomena in large systems, we are especially concerned about how the accuracy scales as system size grows. Therefore, we have also carried out benchmarking of the TW simulations with regard to the scaling of error as system size increases, to the degree possible on small systems. 

To do so, we need to move to Fock state initial conditions as the Hilbert space dimension for open systems with more than 2 sites and coherent state initial conditions with mean occupations $n_b\gg1$ becomes very difficult to deal with. Even though a QT simulation of even $4\times 1$ or $2\times2$ coherent systems is still marginally doable with big numerical effort, the results are too noisy for a good benchmarking. Moving to Fock state incoherent initial conditions gives a Hilbert space that is significantly smaller because of the lack of high occupation tails. Therefore it allows a benchmark with respect to system size to be carried out. 

The Fock-TW simulations implemented as described in Sec.~\ifarxiv{S1 F}{\ref{S-TW-Fock}} were compared with QT quasi-exact simulations also starting from the Fock initial state. 
The mismatch in both the cavity occupation $N_a$ and the phase correlation $g(1)$ are shown in  Fig.~\ref{fig:scalingTW-QT}. One can see that the accuracy improves strongly both as $n_b$ grows but most importantly also as the number of sites grows. The $n_b=3$ data in Fig.~\ref{fig:scalingTW-QT}a also shows that the trend occurs regardless of coherent or Fock initial conditions. 
We also see that for the $n_b=5$ systems studied in the main text, very good accuracy is already obtained for 4 sites ($N_a$) and 6 sites ($g1$).
We emphasize that for $3\times 2$ lattices we could only simulate up to  $N_b=18$ bosons
due to the fast growth of the Hilbert space dimension for the matter component (the total number of Fock states $|n_1,n_2,n_3,n_4,n_5,n_6\rangle$  with $n_i=0,1,..,N_b$ and satisfying $\sum_i n_i=N_b$ is $33649$ for $N_b=18$).

\begin{figure}[htb]
\includegraphics[width=\columnwidth]{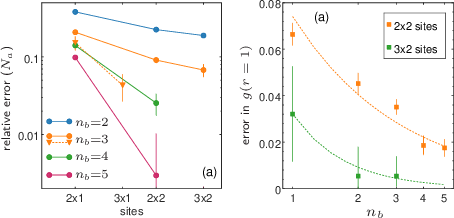}
\caption{
Scaling of the accuracy as system size and $n_b$ increases. 
Panel (a) shows the relative error  $|N_a^{TW}-N_a^{QT}|/N_a^{QT}$ in the cavity occupation $N_a$ for several small systems with both QT and TW evolution starting from the Fock state (round points with solid lines). 
Triangles with a dashed line show the same for $n_b=3$ coherent state initial conditions.
Panel (b) shows absolute discrepancy in the correlation function $g(r=1)$. The dashed lines are power law fits $0.074n_b^{-0.87}$ and $0.032n_b^{-1.8}$ for $2\times 2$ and $3\times 2$, respectively.
Other model parameters are as in Fig. 1c of the manuscript.
Times are $t=1000J$ except for the few largest systems. The $n_b=1$, $3\times2$ case is at $t=500J$ while $t=100J$ is used for $3\times 2, n_b=2,3$ and the $3\times 1$ coherent run.
\label{fig:scalingTW-QT}}
\end{figure}

The very advantageous scaling with system size can be attributed to the relative reduction in size of the discarded third order derivatives in the FPE (\ref{fpe}). To study this in a broad way we compare the estimated size of the discarded third order terms to the kept terms containing the same physics. 

First consider the interaction $U$, and a distribution $W(\alpha,\vec{\beta})$ of widths $\sigma_{\alpha}$ and $\sigma_{\beta_j}$ and peak height $W_0$. The magnitude of the first order term in $U$ for the $\beta_j$ evolution can be estimated as $F^{(1)}_{U,j} \sim  U|n_j-1|\sqrt{n_j} W_0/\sigma_{\beta_j}$, where $n_j$ is the mean occupation of the $j$th lattice site. Here we take $\beta_j\sim\sqrt{n_j}$, $\partial W/\partial\beta_j\sim W_0/\sigma_{\beta_j}$ and the first term in the chain rule expansion of $(\partial/\partial\beta_j)(A_j W) = \partial A_j/\partial\beta_j + A_j\partial W/\partial\beta_j$ is ignored because all constant terms in the FPE cancel to zero.
The magnitude of the 3rd order term is estimated similarly as
$F^{(3)}_{U,j}\sim (UW_0/2\sigma_{\beta_j}^2)[2+\sqrt{n_j}/\sigma_{\beta_j}]$. 
Therefore the ratio of discarded to kept terms is
\eq{Ut3}{
\frac{F_{U,j}^{(3)}}{F_{U,j}^{(1)}} \sim \frac{2+\sqrt{n_j}/\sigma_{\beta_j}}{2\sigma_{\beta_j}|n_j-1|\sqrt{n_j}}. 
}
 $\sigma_{\beta_j}$ can be estimated  for some typical states as follows: coherent states \eqref{ictw} have $\sigma=1/\sqrt{2}$, thermal states in a mode with energy $\ve$ have $\sigma=\sqrt{\tfrac{1}{2}+k_BT/\ve}$, number states in a single mode with occupation $n\gg1$: $\sigma\approx1/4\sqrt{n}$. Such number states in particular modes are not expected to occur in long time states, though, see e.g. Fig.~\ref{fig:wigner-nb}. Taking then as a typical case $\sigma_{\beta_j}=1$, with $n_j\sim n_b$ and significant occupations $n_b\gg1$, one obtains the scaling estimate 
\eq{Ut3fin}{
\frac{F_{U,j}^{(3)}}{F_{U,j}^{(1)}}  \sim \frac{1}{2n_b}.
}
We recover then the well known result that the Wigner truncation becomes negligible for simulating inter-particle interaction $U$ when mode occupations $n_b$ are high. 

Let us look now in the same way at the now well-known truncation of the cavity-lattice interaction $\Omega$.
The truncated terms $F_{\Omega,j}^{(3)}=i(\Omega f_j/4\sqrt{N})\partial^3W/\partial\beta_j\partial\beta_j^*\partial\alpha$ need to be compared to the drift terms for both $\beta_j$ and $\alpha$. Proceeding as before, first for the drift term for $\beta_j$, taking $\alpha\sim\sqrt{N_a}$,  one finds
\eq{Ot3}{
\frac{F_{\Omega,j}^{(3)}}{F_{\Omega,j}^{(1)}} \sim \frac{1}{8\sigma_{\beta_j}^2\sigma_{\alpha}\sqrt{N_a}(1+\sqrt{n_j}/\sigma_{\beta_j})}
}
For the $\alpha$ drift term, we first recognize $\sum_jf_j(|\beta_j|^2-\tfrac{1}{2})$ as $\Delta$, the checkerboard order parameter (n.b. since $f_j$ is staggered, $\sum_j\tfrac{1}{2}f_j\approx 0$). The ratio estimate then comes out as:
\eq{Ot3a}{
\frac{F_{\Omega,a}^{(3)}}{F_{\Omega,a}^{(1)}} \sim \frac{1}{4\sigma_{\beta_j}^2\Delta}
}
Estimating the $\sigma$'s as 1 and $n_b\gg1$ as before we obtain the scaling estimates
\eq{Ot3fin0}{
\frac{F_{\Omega,j}^{(3)}}{F_{\Omega,j}^{(1)}} \sim \frac{1}{8\sqrt{N_an_b}}\quad;\quad
\frac{F_{\Omega,a}^{(3)}}{F_{\Omega,a}^{(1)}} \sim \frac{1}{4\Delta}.
}
Finally, in phases with nonzero $\Delta$, we see e.g. in Fig.~2 of the main text or Fig.~\ref{fig:bc-ad} that $\Delta\sim\mathcal{O}(N)$, while it is seen from the adiabatic approximation \eqref{adiab} that $N_a\approx a_{\rm est}^2 = \Omega^2 \Delta^2/(N(\delta^2+\kappa^2))$, and therefore $N_a\sim \Omega^2N\times$const.
Noting that $N=n_b M$, we then obtain final scaling estimates for single site quantities and cavity quantities, respectively, as 
\eq{Ot3fin}{
\frac{F_{\Omega,j}^{(3)}}{F_{\Omega,j}^{(1)}} \sim \frac{1}{8\Omega n_b\sqrt{M}}\quad;\quad
\frac{F_{\Omega,a}^{(3)}}{F_{\Omega,a}^{(1)}} \sim \frac{1}{4n_b M},
}
in terms of mean lattice occupations $n_b$ and system size in terms of lattice sites $M$.

Comparing to \eqref{Ut3fin}, the cavity-related discarded contributions scale more rapidly by $1/\sqrt{M}$ and $1/M$, \emph{showing that these corrections are always sub-leading as compared to $n_b^{-1}$, provided the system is large enough.}

From the above we see overall that accuracy of the simulation is expected to improve rapidly approximately as $1/n_b$ when $n_b\gtrsim 1$, and the accuracy of physics related to the cavity-lattice coupling $\Omega$ is further expected to improve proportionally to $1/M$ and $1/\sqrt{M}$. The scaling of accuracy with $M$ and $n_b$ seen in Fig.~\ref{fig:scaling} appears to be roughly consistent with the above, modulo statistical errors and likely finite-size issues for such small systems.

\subsection{F. Number state initial conditions}
\label{S-TW-Fock}

If needed, a convenient and scalable TW representation of approximate Fock states is available via the prescription found in  \cite{Olsen09}. 
The exact Fock-state Wigner function is 
\eq{Wfull}{
W(\beta) = \frac{2(-1)^n}{\pi} e^{-2|\beta|^2}L_N(4|\beta|^2),
}
where $L_N$ are the Laguerre polynomials of order $N$. One can, in principle, map to stochastic variables directly from \eqref{Wfull} with the use of positive/negative weights, but such an approach suffers from a sign problem and is therefore unusable for large systems.

Instead, the prescription for a Fock state with occupation $n$ that is scalable to large systems was derived in \cite{Olsen09} and is
\eq{ictwfock}{
\beta_j(0) =  e^{2i\pi u_j}\left(R_n+\frac{\zeta_j}{4R_n}\right),
}
where $\zeta_j$ are independent Gaussian random variables with mean zero, variance $1$, $u_j$  are uniformly distributed random numbers in the interval $[0, 1)$, and the optimal radius in phase space is 
\eq{ictwfockp}{
R_n = \frac{1}{2}\sqrt{2n+1+2\sqrt{n^2+n}}.
}
It converges to $R_n\approx\sqrt{n}$ as $n$ grows.

The prescription \eqref{ictwfock} is what we have used to implement Fock state initial conditions using $n=n_b$ and a vacuum state in the cavity mode $\alpha$. 
What is omitted in the above 
are the positive/negative value oscillations that appear far from the dominant value of $|\beta|=\sqrt{n}$. These oscillations become small as $n_b$ grows, so concurrently the prescription converges rapidly to the exact Wigner representation.

\begin{figure}[htb]
\includegraphics[width=\columnwidth]{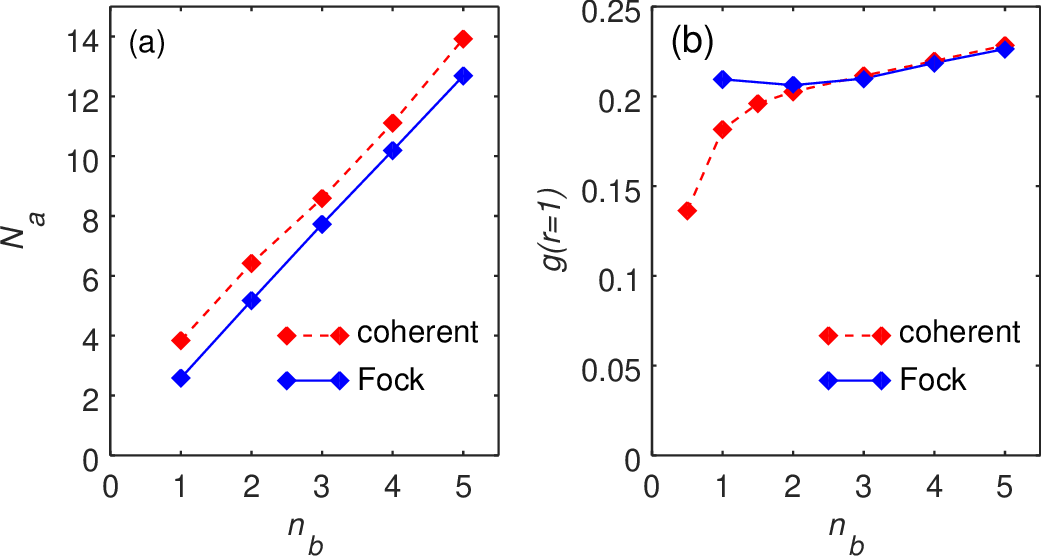}
\caption{
Effect of coherent state or Fock number state initial conditions on long-time values ($tJ=1000$) of cavity occupation $N_a$ and correlation $g(1)$ for the $2\times1$ system.
The difference stems from the conserved distribution (variance) of total boson number $N$.
All values are from QT simulations.
Parameters  are as in Fig. 1c of the manuscript apart from the mean occupation per lattice site $n_b$, which we vary in this figure.
\label{fig:ic}}
\end{figure}

 As the boson number on the lattice  is conserved by our Hamiltonian and dissipation, there are small quantitative differences between the stationary states resulting from Fock and coherent state initial conditions (see Fig.~\ref{fig:ic}), even when the mean boson number is the same. This is because the coherent state contains a spread of boson number, and high $N_b$ realizations lead to higher cavity occupations on average, in a nonlinear way. The difference in $N_a$ at long times are about 10\% for $n_b=5$, while the difference in $g(1)$ is negligible for any $n_b>2$.

\subsection{G. Operational indicators of TW accuracy}
\label{S-TW-acc}

The conditions under which the third order terms in the Fokker-Planck Equation \eqn{fpe} can be omitted have been benchmarked in detail above, but have also been the focus of quite a number of investigations, e.g. \cite{Sinatra02,Norrie06,Kinsler91,Drobny97,Plimak01,Polkovnikov03,Deuar07}, especially in the context of the simulation of quantum optical and ultracold atom systems with interaction $U$. In these studies,
several practical operational conditions for a reliable simulation have emerged 
\cite{Blakie08,Sinatra02,Norrie06}:
\begin{enumerate}
    \item The mean occupation per mode should be greater than $\approx\tfrac{1}{2}$ \cite{Sinatra02,Norrie06}. However, notably, single mode occupations need not meet the criterion provided the overall system does \cite{Norrie06}.
    \item Local observables such as $n_j$ and the density-density correlations $g^{(2)}(r)=\langle\op{n}_0(\op{n}_r-\delta_{0r})\rangle/(n_0n_r)$ should remain within physical limits such as $n_j>0$ and $g^{(2)}(r)>0$.
    \item For Hamiltonian systems that redistribute mode occupations under internal Hamiltonian dynamics, problems consisting of incorrect redistribution have been seen when the maximum excitation energies allowed exceed several times $k_BT_{\rm eff}$. Here $T_{\rm eff}$ is the final effective temperature reached at long times \cite{Sinatra02}. This condition is correlated with condition 1, in that allowing very many high energy modes above thermal energies, so that they are poorly occupied, decreases the mean occupation per mode.
\end{enumerate}

\begin{figure}[htb]
\includegraphics[width=\columnwidth]{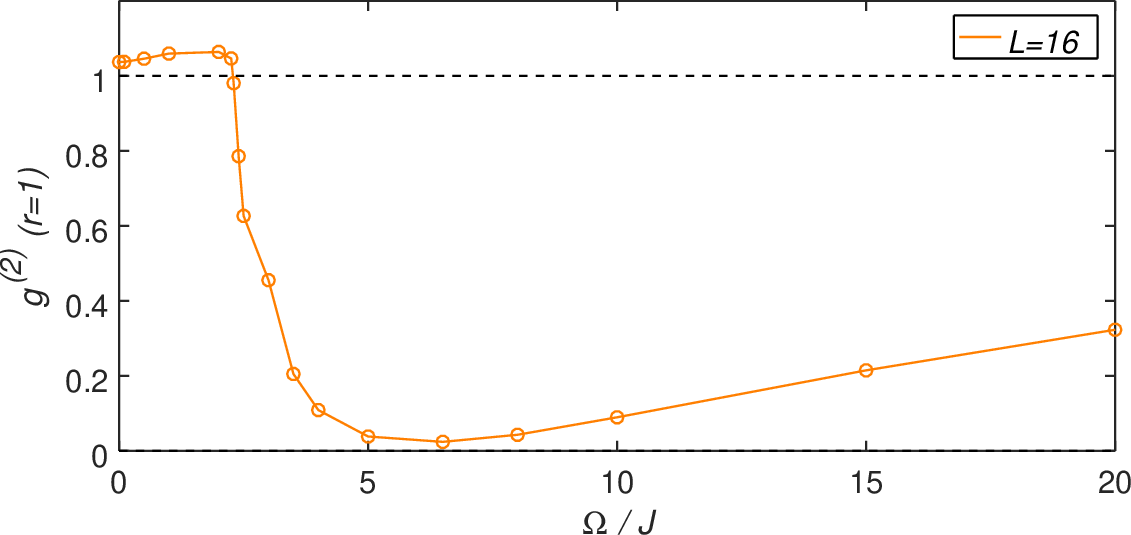}
\caption{Nearest neighbor density-density correlation $g^{(2)}(r=1)=\langle\dagop{b}_0\dagop{b}_1\op{b}_0\op{b}_1\rangle/n_b^2$ in the long-time state at $tJ=10^4$ as a function of the cavity coupling $\Omega$ for a $16\times16$ lattice.
\label{fig:wigner-g2}}
\end{figure}

\begin{figure}[htb]
\includegraphics[width=\columnwidth]{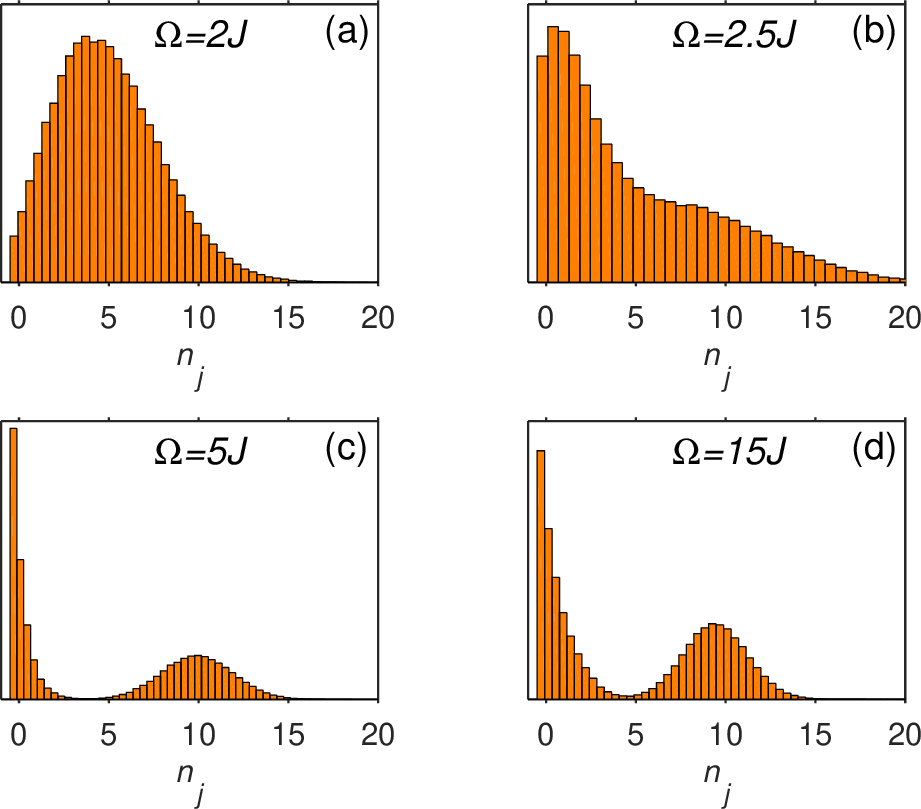}
\caption{Distribution of site occupations $n_j=|\beta_j|^2-\tfrac{1}{2}$ at $tJ=10^4$, $L=16$ for several characteristic $\Omega$  values. 800 trajectories.
\label{fig:wigner-nb}}
\end{figure}

How do our system and simulations compare to the above criteria?

(1) \tick The mean occupation is $n_b=5$, meeting the average occupation criterion well. 

(2) With regard to local observables,
\begin{itemize}
    \item \tick mean site occupations $n_j$ are $n_b=5$ so there are no problems with negative densities on average. Note that some $n_j<0$ in particular trajectories are expected, since they are essential to describe a vacuum.
\item  \tick  With regard to density-density correlation, the lowest values in our system occur for nearest neighbors ($r=1$). 
    Fig.~\ref{fig:wigner-g2} shows the dependence of this correlation $g^{(2)}(r=1)$ on $\Omega$. There is no indication of nonphysical $g^{(2)}<0$ occurring here, though the exact values of $g^{(2)}(r)$ values very close to zero should be taken with a grain of salt.
    \item The distribution of site occupations $n_j$ is shown in Fig.~\ref{fig:wigner-nb} in the main regimes studied. In all cases the overwhelming majority of modes have occupations above 1, and most sites in the low density part of the checkerboard pattern also keep to $n_j>1$. Single shot values are not expected to conform to $n_j>0$ anyhow, since single trajectory values of such low occupations do not correspond precisely to measurements and e.g. a vacuum has half shots with $n_j<0$ and is represented fine. \tick
\end{itemize}
    Overall we see no indications of problems with observables keeping to physical limits.

(3) While temperature is not easy to quantify for our system, we can consider single particle energy, and whether there are many unoccupied high energy modes. As best seen perhaps by comparing \eqn{ssk} to free space, which would have the exponential factor exp[$-i(dt/2)(\hbar k)^2/2m$], the  basic single-particle energies are $\epsilon_{\vec{k}}=2J(2-\cos(2\pi k_1/L)-\cos(2\pi k_2/L)\,)$. The charge density wave, which is significantly occupied for all regimes except for the SF, occupies in fact the highest energy kinetic excitations at $k_1=k_2=L/2$. Therefore the situation of many unoccupied higher energy modes does not take place there. Moreover, our system is not strictly Hamiltonian, therefore one does not expect redistribution problems in modes whose dissipation timescale is shorter than the rather long thermalization time. \tick

Summarizing the checks of the above indicators, the TW simulations meet the conditions considered necessary for good accuracy, and provide for the inclusion of spontaneous and quantum fluctuating effects at least to leading orders, and in a non-perturbative way. 
There are also no direct indications of any appreciable inaccuracies, though some care should be taken with results that depend primarily on low occupied modes.

\begin{figure}[htb]
\includegraphics[width=0.8\columnwidth]{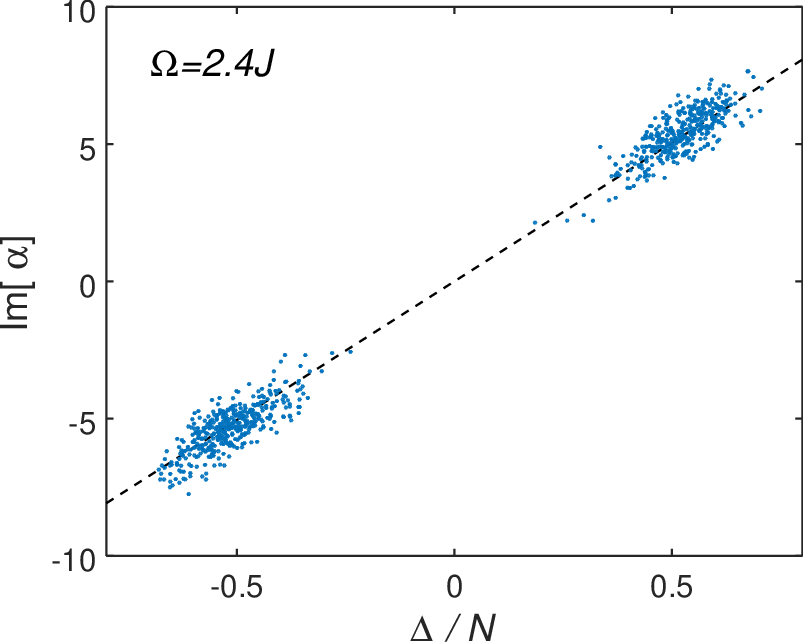}
\caption{Data from 800 single trajectories of the $L=16$ full model: Im[$\alpha$] vs. $\Delta$ (points) at $tJ=10^4$, compared to the adiabatic elimination estimate \eqn{adiab} (dashed line).
\label{fig:bc-ad}}
\end{figure}

\begin{figure}[htb]
\includegraphics[width=0.49\columnwidth]{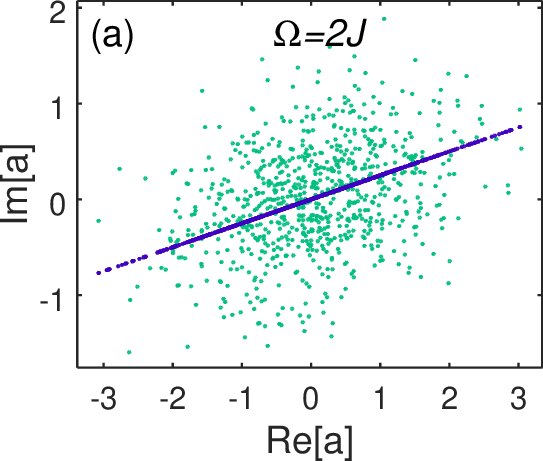}
\includegraphics[width=0.49\columnwidth]{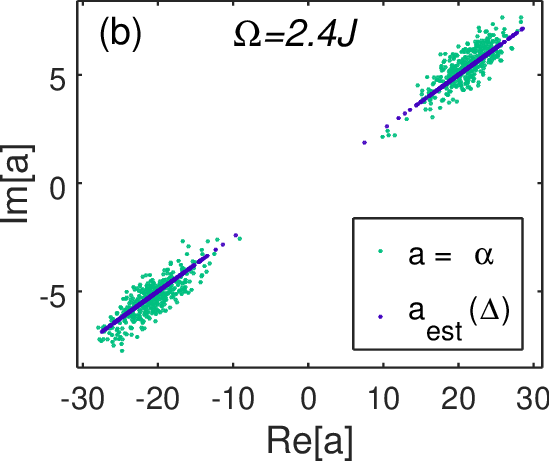}
\setlength{\unitlength}{1mm}
\begin{picture}(80,72)(3.5,-2)
\put(0,0){\includegraphics[width=\columnwidth]{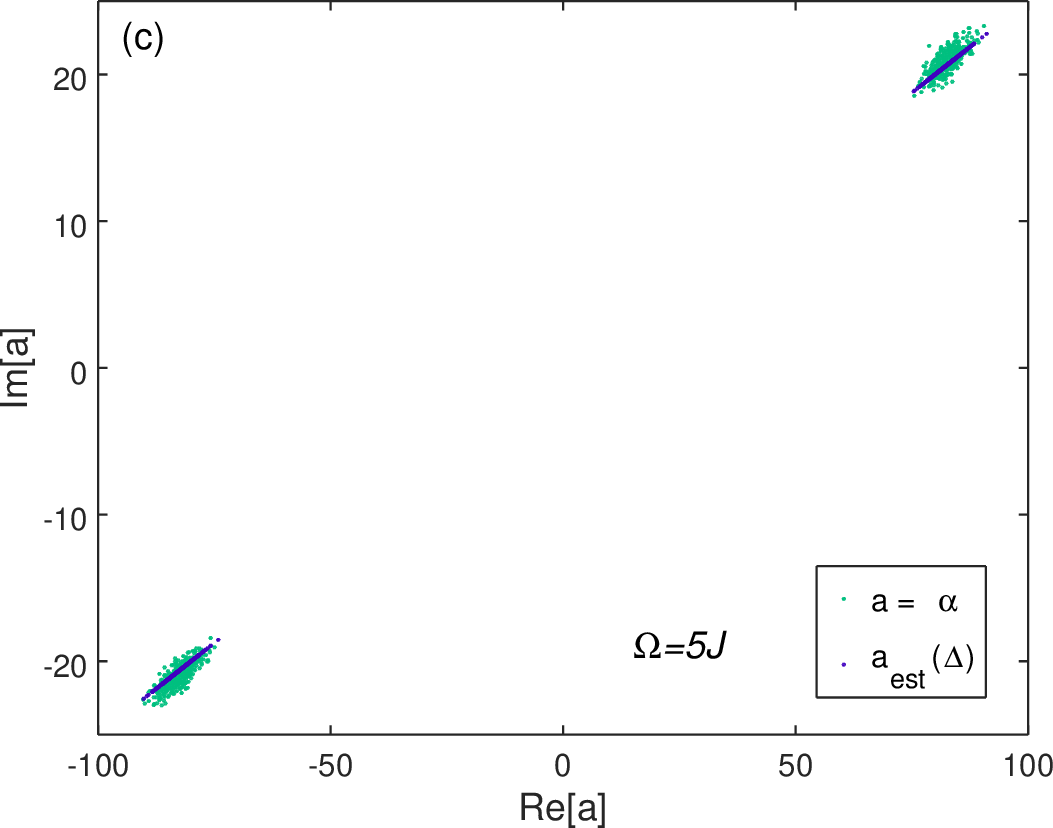}}
\put(15,28){\includegraphics[width=0.54\columnwidth]{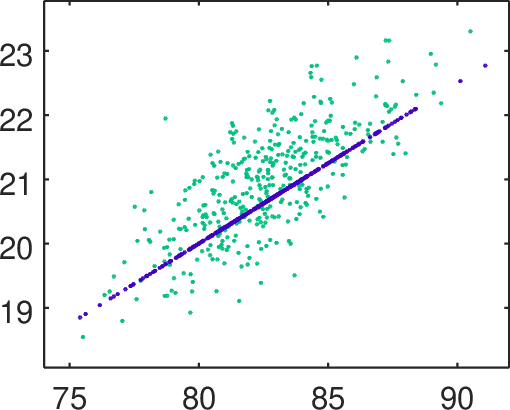}}
\end{picture}
\includegraphics[width=0.4\columnwidth]{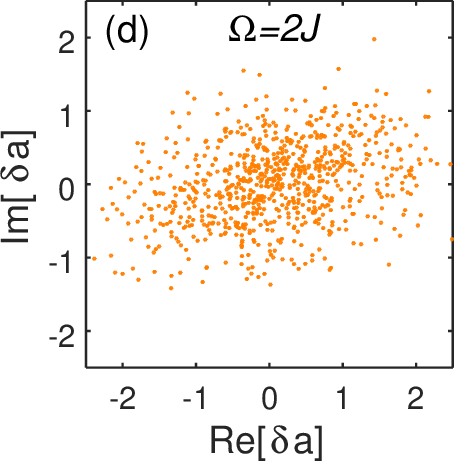}
\hspace*{0.13\columnwidth}
\includegraphics[width=0.4\columnwidth]{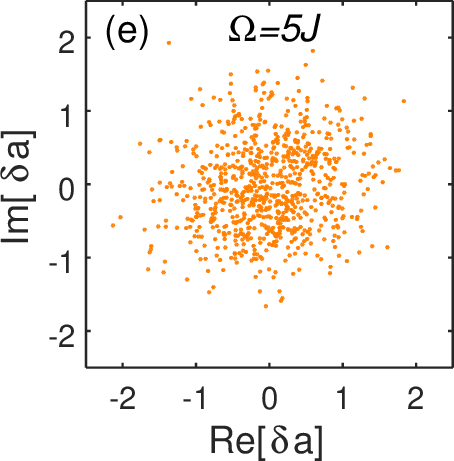}
\caption{Comparison of the actual cavity mode amplitude (turqoise) to that assumed by the adiabatic elimination (``bad cavity'') approximation as given by \eqn{adiab} (dark blue). At $t=10^4$ $L=16$ for several characteristic $\Omega$ in the superfluid (a) and supersolid (b-c) phases. Panels (d-e) show the difference $\delta a=\alpha-a_{\rm est}$ in individual trajectories.
\label{fig:badcavity}}
\end{figure}


\section{S2. Accuracy of the bad cavity approximation}
\label{S-BAD}

The bad cavity (adiabatic elimination) approximation assumes that the dissipative timescale is short enough for the lossy mode (here $\op{a}$) to collapse onto its steady state faster than other system quantities evolve. Therefore it takes the steady state value of its evolution equation when other quantities (here $\beta_j, \Delta$) are assumed constant. 
Taking the steady state limit of \ifarxiv{\eqn{Eq:TW}}{(4)} 
for the cavity mode by imposing $dW=d\alpha=0$, one obtains
$(\delta -i \kappa)\alpha -\frac{\Omega}{\sqrt N}\sum_j f_j\left(|\beta_j|^2-\frac{1}{2}\right)=0$. 
 Solving for $\alpha$ gives the estimate $\alpha\to a_{\rm est}$ where
\eq{adiab}{
a_{\rm est} = \frac{\Omega}{\sqrt{N}}\left(\frac{\Delta}{\delta-i\kappa}\right).
}
Substitution of $\alpha=a_{\rm est}$ into the evolution equations \ifarxiv{\eqn{Eq:TW}}{(4)} for $\beta_j$ gives the TW equations for the EBH model. The EBH model in its operator form is obtained by making the substitution 
\eq{adiabop}{
\op{a}\to \frac{\Omega}{\sqrt{N}}\left(\frac{\op{\Delta}}{\delta-i\kappa}\right)
}
in the Hamiltonian $\hat H$ and by discarding the Linblad term in  the evolution equation \ifarxiv{\eqn{master}}{(3)}.
This gives
\eq{ebhm}{
\op{H}_{bc}+\op{H}_c = -\frac{\Omega^2}{N}\,\frac{\delta}{\delta^2+\kappa^2}\,\op{\Delta}^2,
}
which is a long-distance interaction
since $\op{\Delta}^2=\sum_{ij}f_if_j\op{n}_i\op{n}_j$.

Fig.~\ref{fig:bc-ad} shows the scatter of the cavity mode $\alpha$ in single realizations with respect to the one-to-one relationship \eqn{adiab} indicated in the bad cavity limit.
Fig.~\ref{fig:badcavity} shows the relationship in more detail -- for each realization it compares the approximation \eqn{adiab} made when using the atomic $\Delta$ with the actual cavity mode amplitude.  
A random valued discrepancy is evident in both regimes: superfluid (panels a,d) and supersolid (panels b-c,e). It is particularly strong in the super-fluid, while in the density correlated phases the adiabatic approximation $a_{\rm est}$ retains the overall sign and approximate phase seen in the full model, but collapses the photon amplitude to a single line. The error in $\alpha$ is of the order of $\mc{O}(1)$ for our parameters, with a more or less random phase. It corresponds to an error of order $\sqrt{N_a}$ in the cavity photon number and of order $1/\sqrt{N_a}$ in the cavity phase.

\section{S3. Details of the Dicke SF / SS transition}
\label{S-Dicke}

First, Fig.~\ref{fig:SpiNa} shows the very close relationship between the static structure factor and the cavity mode occupation. In particular 
$S_{\pi,\pi}/(n_bL)^2 \propto N_a/\Omega^2$ is quite closely followed in the whole range of $\Omega$ we have studied.

\begin{figure}
\vspace{0.5cm} 
\includegraphics[width=\columnwidth]{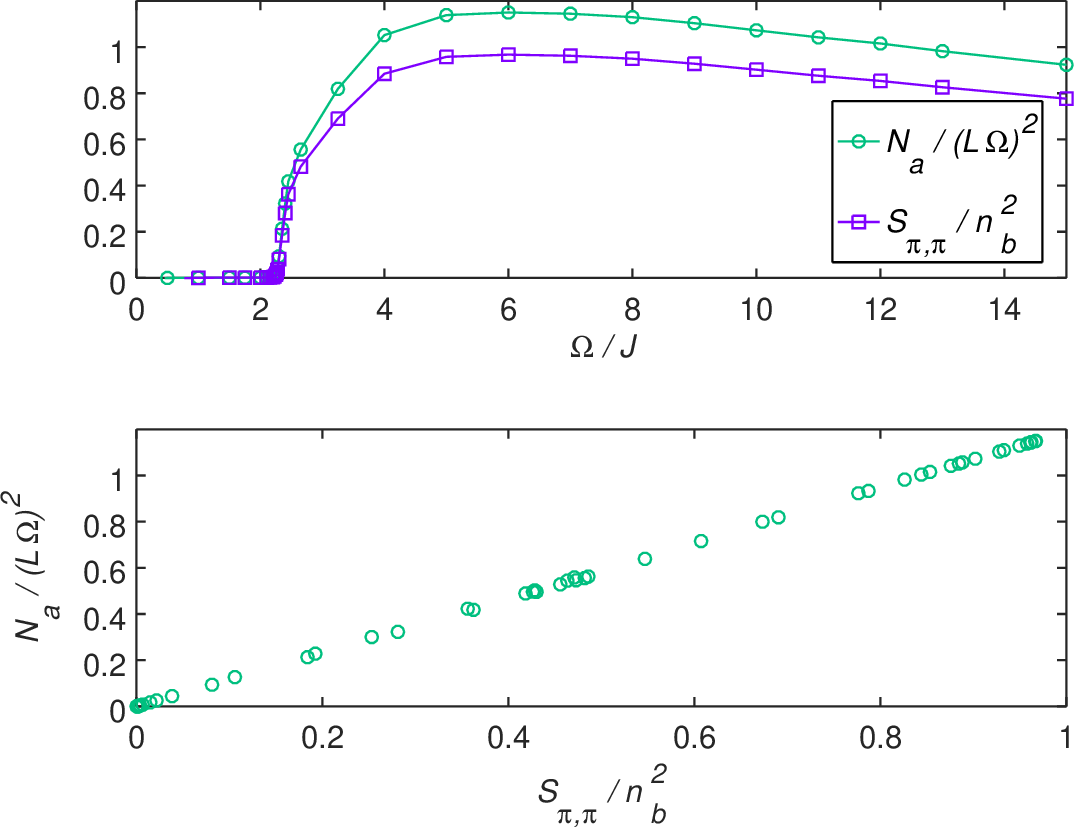}
\caption{Behavior of the static structure factor $S_{\pi,\pi}$ and cavity occupation $N_a$ and the relationship between them. Shown for the system with the usual parameters at $tJ=10^4$, on a $16\times16$ lattice, averaged over 800 trajectories.
\label{fig:SpiNa}}
\end{figure}

Figure.~\ref{fig:scaling} shows the convergence of finite-size values of $S_{\pi,\pi}$ as $L$ grows. The plot is tailored to a $1/L^2$ scaling, according to  \ifarxiv{\eqn{SpiL2}}{(6)}  
which -- if correct -- would correspond to a linear behavior of data in Fig.~\ref{fig:scaling}. Looking at the data, the trend holds well at large $L$, trending slightly down (i.e. to a smaller power of $1/L$) at small system sizes on the right of the plot. As expected from critical coarsening, this trend is most pronounced for $\Omega$ values closest to the transition point. There a given $L$ is effectively further from the thermodynamic limit than other $\Omega$ values.
Fig.~\ref{fig:scalefit}, shows the results of fits to the ansatz~(6).

\begin{figure}[tb]
\includegraphics[width=\columnwidth]{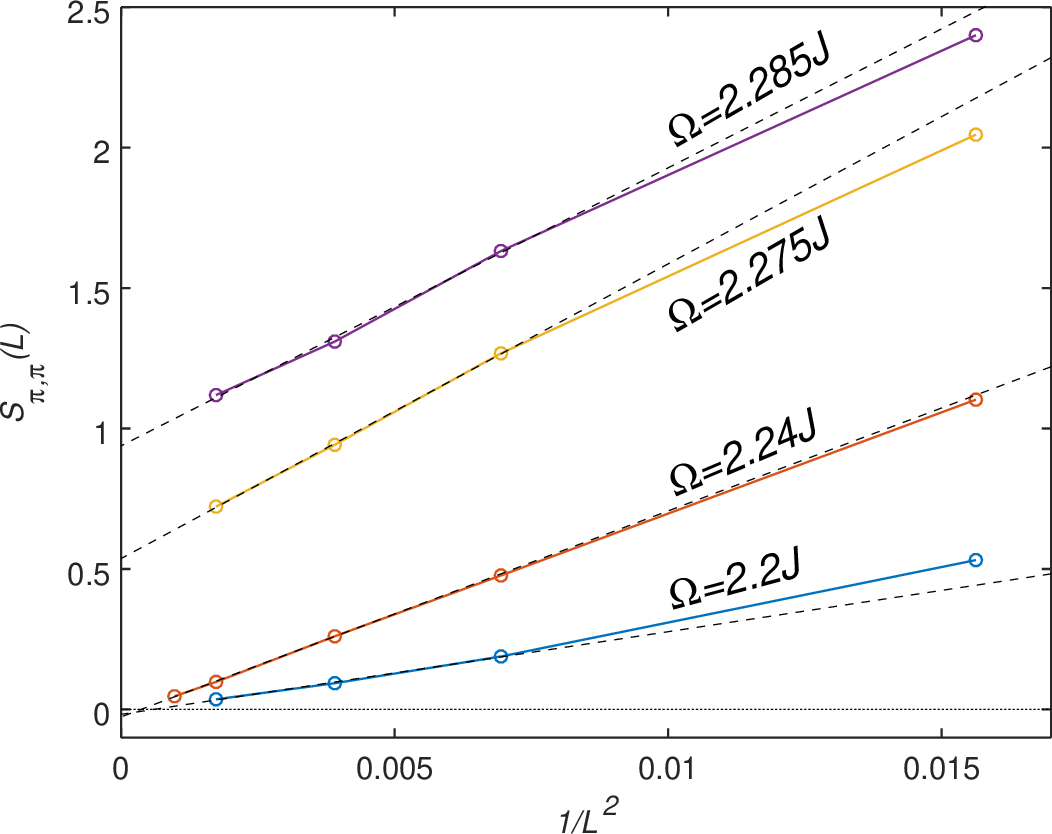}
\caption{Finite size scaling towards $L\to\infty$ for the $S_{\pi,\pi}$ observable at several values of $\Omega$ near the Dicke SS/SF phase transition. Circles are simulation data with connecting colored lines. The black dashed lines are linear fits to the three highest $L$ values. Parameters and data are the same as shown in Fig.~\protect\ifarxiv{\ref{Fig:Dicke}}{4}. 
\label{fig:scaling}}
\end{figure}

The growth of the density modulation order parameter (static structure factor $S_{\pi,\pi}$) is very well described as a linear growth at rate $G$ above the transition point:
\eq{lingrow}{
S_{\pi,\pi} = \left\{
\begin{array}{c@{\qquad\text{if}\ }l}
  G(\Omega-\Omega_s)   &  \Omega>\Omega_s\\
  0   & \Omega\le\Omega_s
\end{array}
\right.
}
This is shown in Fig.~\ref{fig:scalefit}a, along with a fit. The fit gives $G=51.3$, $\Omega_s=2.265$. An error estimate can be obtained from the difference when compared to a fit over only the four rightmost data points $\Omega\ge2.285$ instead of five. Taking twice this difference as the error estimate, one arrives at
\eq{Oms-lin}{
\Omega_s = 2.265(5)\qquad;\qquad
G= 51.3(9).
}
The scale coefficient $C\Omega)$  is peaked around the transition point as expected.

\begin{figure}[htb]
\includegraphics[width=\columnwidth]{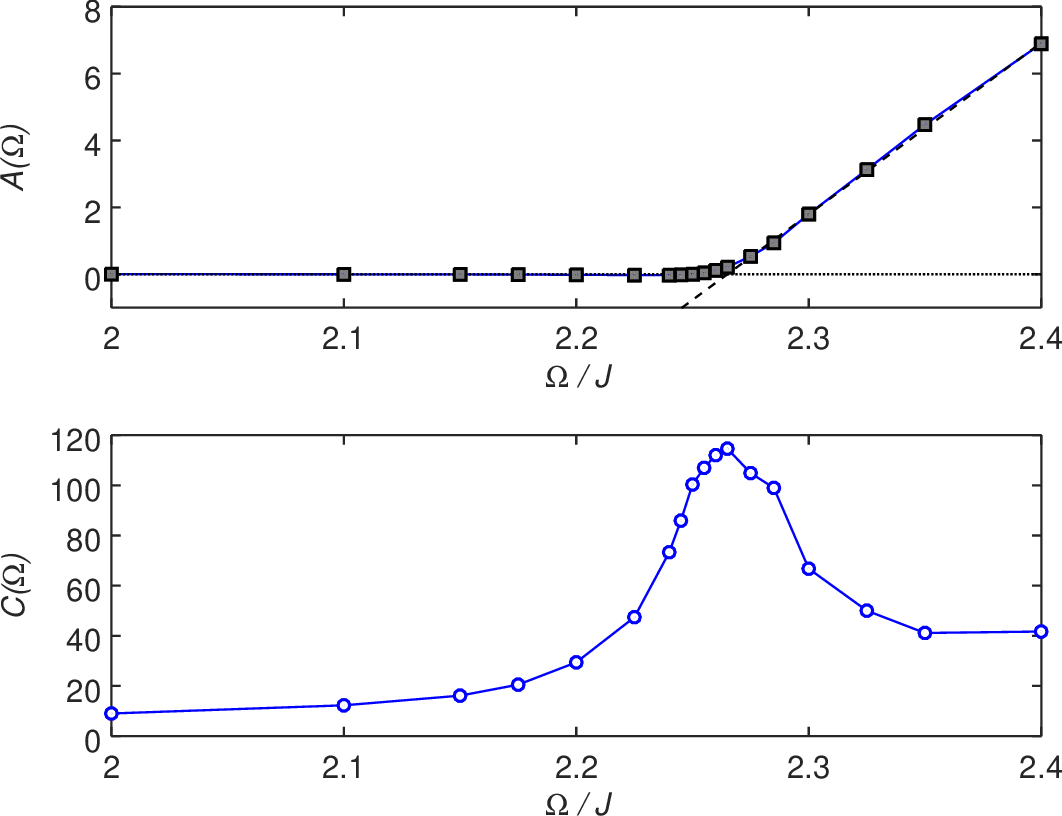}
\caption{Fitted parameters to Eq.~\protect\ifarxiv{\eqn{SpiL2}}{(6)}
around the 
SF/SS phase transition. Top: $A(\Omega)$, the thermodynamic limit of $S_{\pi,\pi}$, as grey squares with interpolation between as a blue line. The dashed line is a fit to \eqn{lingrow} using the rightmost five data points $\Omega\ge2.275$.
The bottom panel plots the finite scale susceptibility $C(\Omega)$.
\label{fig:scalefit}}
\end{figure}

Details of the behavior of the magnetization $\Delta$ across the transition point are shown in Fig.~\ref{fig:binder-detail}. The distribution in panel b at $\Omega=2.26$, just below the transition corresponds to negative Binder cumulants. The cause of $U_L<0$ here are the fatter tails compared to the Gaussian distribution seen at lower $\Omega=2.25$. This means that a few realizations are already achieving non-negligible ordering $|\Delta|\gtrsim0.2N$. The panels c-d which proceed across the transition show a broadening of the distribution to include successively more ordered realizations with $|\Delta|>0.2N$ but still without the exclusion of intermediate unordered cases that is only seen once the transition has been unambiguously crossed at $\Omega=2.325$ in panel e.

\begin{figure}
\vspace{0.5cm} 
\includegraphics[width=\columnwidth]{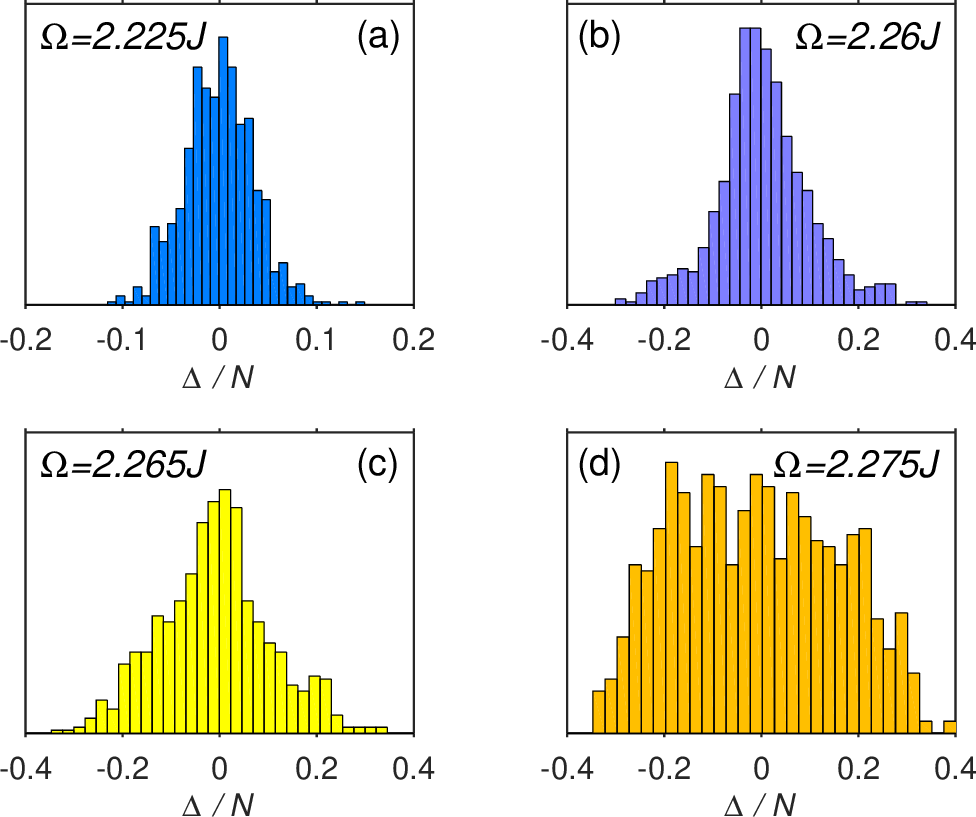}\\
\includegraphics[width=0.445\columnwidth]{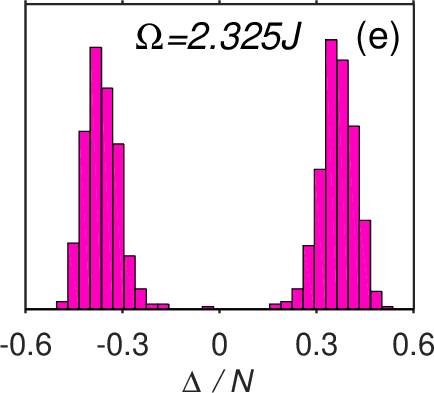}\hspace*{0.1cm}
\includegraphics[width=0.515\columnwidth]{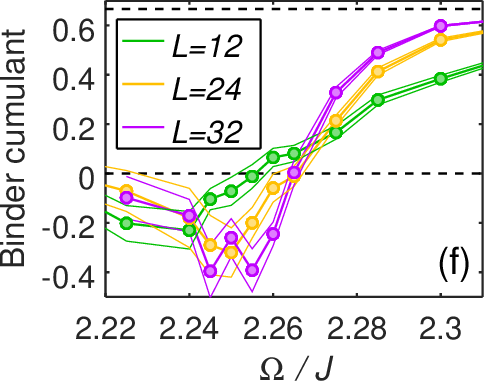}\hspace*{0.12cm}
\caption{Behavior of the $\Delta$ effective magnetization near the Dicke transition. (a-e) show the $\Delta$ distributions for $L=32$ around the transition, while (f) shows detail of the Binder cumulant $U_L$ in the vicinity. 
\label{fig:binder-detail}}
\end{figure}

\section{S4. Condensate fraction near the BKT phase transition}
\label{S-Cond}
The condensate fraction is defined as the ratio $N_0/N$, where
$N_{k=0}=\sum_r \langle b^\dagger_0 b_r \rangle $ 
is the number of bosons in the zero momentum state, forming the condensate.  It is the  analogue of the susceptibility in spin models and has been measured in current experiments with cold atoms in optical cavities \cite{Landig2016}.
In Fig.~\ref{Fig:CondensateFraction}  we display the behavior of  
the condensate fraction at long times in the vicinity of the BKT phase transition as a function of the Rabi frequency for different system sizes. Differently from the correlation ratio, the condensate fraction scales to zero with system size
\emph{even} in the QLRO phase. This is consistent with the Hohemberg-Mermin-Wagner theorem, stating that no true condensation can exist in the thermodynamic limit for 2D systems with continuous U(1) symmetry. This result is well known for models at thermal equilibrium, like the XY model or the Bose-Hubbard model, but it has also been verified for non-equilibrium 2D systems, like driven-dissipative photon-polaritons \cite{DagvadorjPRX2015}. 

\begin{figure}
\includegraphics[width=1\columnwidth,trim={2.5mm 2mm 16mm 14.5mm},clip]{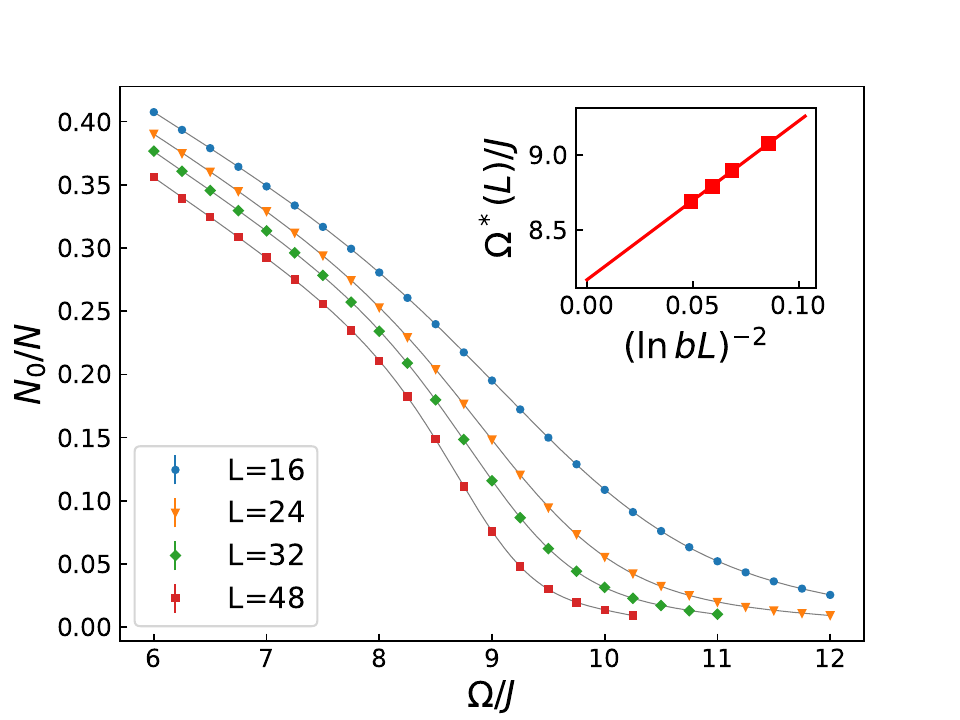}
\caption{Main panel: condensate fraction $N_0/N$ as a function of the Rabi frequency for increasing system sizes.  The solid lines are Padé approximant fits to the numerical data that we use to extract the 
 derivative of the data curve, from which we extract inflection points $\Omega^*(L)$. Inset: scaling behavior of the inflection points based on Eq.~\protect\ifarxiv{\eqn{Eq:FSS}}{(7)}, 
 with $\Omega_{KT}(L)$ replaced by  $\Omega^*(L)$, yielding $\Omega_c/J=8.17(14)$.
}
\label{Fig:CondensateFraction}
\end{figure}

Importantly, we see from Fig.~\ref{Fig:CondensateFraction} that the data curves  become steeper and steeper in the vicinity of $\Omega\approx8J-9J$ as the system size increases. In the thermodynamic limit the condensate fraction would vanish, with an infinite derivative at the BKT critical point.  
As previously shown for the 2D Bose-Hubbard model at finite temperature \cite{CarrasquillaPRA2012}, one can approximately estimate the position of the critical point by 
studying the scaling behavior of the inflection points $\Omega^*(L)$ in the data curves of Fig.~\ref{Fig:CondensateFraction}.
The correlation length $\xi$ at the inflection point must be 
of the order of the system size, that is $\xi(\Omega=\Omega^*(L))=bL$, where $b$ is a constant of order unity. By replacing $\xi$ with the BKT form
$\xi=\exp(c/\sqrt {\Omega/\Omega_{c}-1})$,
we recover Eq.~\ifarxiv{\eqn{Eq:FSS}}{(7)}
with $\Omega_{KT}(L)$ replaced by $\Omega^*(L)$. We can therefore extrapolate the critical Rabi frequency by fitting the 
inflection points data based on Eq.~\ifarxiv{\eqn{Eq:FSS}}{(7)}.
The results obtained are shown in the inset of Fig.~\ref{Fig:CondensateFraction}. In particular we find  $\Omega_c/J=8.17(14)$, which is consistent with, but less precise, than the estimate based on $r_L$ reported in the main text.

\end{document}